\documentclass[pra,twocolumn,english,showpacs,floatfix]{revtex4-1}
\usepackage{graphicx}
\usepackage{amssymb}
\usepackage{color}
\usepackage{amsmath}
\usepackage{enumerate}
\usepackage{graphicx}
\newcommand\abs[1]{\left|#1\right|}
\newcommand\aver[1]{\langle#1\rangle}
\newcommand{\beq}{\begin{equation}}
\newcommand{\eeq}{\end{equation}}

\graphicspath{{figures/}}

\begin{document}
\title{Cavity-induced chiral states of fermionic quantum gases}
\author{Ameneh Sheikhan$^1$, Ferdinand Brennecke$^2$, Corinna Kollath$^1$}

\affiliation{$^1$HISKP, University of Bonn, Nussallee 14-16, 53115 Bonn, Germany\\
$^2$ Physikalisches Institut, University of Bonn, Wegelerstr.~8, 53115 Bonn, Germany}

\begin{abstract}
We investigate ultra-cold fermions placed into an optical cavity and subjected to optical lattices which confine the atoms to ladder structures. A transverse running-wave laser beam induces together with the dynamical cavity field a two-photon Raman-assisted tunneling process with spatially dependent phase imprint along the rungs of the ladders. We identify the steady states which can occur by the feedback mechanism between the cavity field and the atoms. We find the spontaneous emergence of a finite cavity field amplitude  which leads to an artificial magnetic field felt by the fermionic atoms. These form a chiral insulating or chiral liquid state carrying a chiral current. We explore the rich state diagram as a function of the power of the transverse laser beam, the atomic filling, and the phase imprint during the cavity-induced tunneling. Both a sudden onset or a slow exponential activation with the transverse laser power of the self-organized chiral states can occur. 
\end{abstract}

\maketitle
\section{Introduction}
The physics of the coupling between atoms and electromagnetic fields has a long history. For example, laser light has been employed,  to cool atoms to previously unreachable temperatures and to trap and manipulate them even at quantum degeneracy \cite{Chu1998,CohenTannoudji1998,Phillips1998, Foot2005, PethickSmith}. Typically, in these situations the back-action of the atoms onto the laser light can be neglected. However, this changes drastically as soon as the quantum nature of the photon field comes into play and the atoms start to interact several times with a single photon. Experimentally this situation can be reached if atoms are placed into an optical cavity \cite{RitschEsslinger2013}. One of the consequences of the presence of atoms is for example a density dependent shift of the cavity resonance frequency.

More recently, the great experimental advances have allowed one to realize the so-called Dicke-phase transition \cite{Dicke1954,HeppLieb1973, WangHioe1973}. To this end, a quantum degenerate bosonic gas was placed into a high-finesse optical cavity subjected to a transverse off-resonant pump beam. Above a critical pump strength, the feedback between the atomic density distribution and the cavity field leads to a spontaneous formation of a symmetry broken phase in which the atoms form a checkerboard density pattern off which pump light is super-radiantly scattered into the cavity \cite{DomokosRitsch2002,NagyDomokos2008,BaumannEsslinger2010,KlinderHemmerich2015,RitschEsslinger2013,PiazzaZwerger2013,DimerCarmichael2007,BadenBarrett2014}. Details of the steady-state diagram as for example different super-radiant fixed points~\cite{BhaseenKeeling2012,LiuJia2011}, dynamic correlations \cite{KulkarniTuereci2013}, the damping of quasi-particles \cite{KonyaDomokos2014}, self-ordered limit cycles \cite{PiazzaRitsch2015}, or prethermalization effects \cite{SchuetzMorigi2014} have been investigated theoretically.

Another exciting situation has been reached experimentally by the additional application of external optical lattice potentials~\cite{KlinderHemmerich2015b,LandingEsslinger2015}. In such a setup, a modified  Bose-Hubbard model of the bosonic quantum gas can be reached and the influence of cavity-induced, long-range interactions between the atoms onto the superfluid to Mott-insulator phase transition has been investigated \cite{LarsonLewenstein2008, MaschlerRitsch2005, MaschlerRitsch2008, NiedenzuRitsch2010,SilverSimons2010,VidalMorigi2010, LiHofstetter2013,RitschEsslinger2013,BakhtiariThorwart2015}.

Theoretically, further proposals have been put forward for the self-organization of complex quantum phases in combined cavity-atom systems. For example, the organization of bosonic atoms into triangular or hexagonal lattices \cite{SafaeiGremaud2015} or of fermionic atoms into super-radiant phases \cite{LarsonLewenstein2008b,MuellerSachdev2012,PiazzaStrack2014,KeelingSimons2014,ChenZhai2014} have been pointed out. In more complex setups such as multi-mode cavities \cite{GopalakrishnanGoldbart2009,NimmrichterArndt2010,StrackSachdev2011,GopalakrishnanGoldbart2011,HabibianMorigi2013,JanotRosenow2013, BuchholdDiehl2013} complex disordered structures, such as glasses or complex supersolids have been proposed. Moreover, phases in which spin-orbit coupling becomes important have been suggested in standing-wave cavities \cite{DengYi2014,DongPu2014,PanGuo2015,PadhiGosh2014} or ring cavities \cite{MivehvarFeder2014,MivehvarFeder2015}. 

Coupled cavity-quantum gas systems not only provide a platform to realize novel self-organized collective phases, but also offer via the cavity output field valuable information about the atomic state in real time and in a non-destructive way. Such measurements have been proposed \cite{MekhovRitsch2007,ChenMeystre2007,ChenMeystre2009,BhattacherjeeMan2010, MekhovRitsch2012,OeztopTuereci2012, KozlowskiMekhov2015} and conducted \cite{Brennecke2013, Landig2015} in order to extract equal or many-time correlation functions of the atomic gas or the atomic quantum statistics.  

The field of cavity physics has very recently been connected to the lively and exciting field of topologically non-trivial quantum phases \cite{PanGuo2015,PanYi2015, KollathBrennecke2016}. 
The interest in the field of topologically non-trivial effects has revived enormously during the last years, in particular, stimulated by the discovery of topologically insulating materials \cite{HasanKane2010}. Topologically non-trivial quantum phases possess special properties such as extended edge modes that can be well protected against destructive environmental effects \cite{HasanKane2010}. Therefore, these materials are promising for technological applications. For example, the utilization of such topologically protected modes lies at the heart of the field of topological quantum computation \cite{SternLindner2013}. 

Topologically non-trivial phases have recently been realized in cold atom experiments using for example strong artificial magnetic fields \cite{DalibardOehberg2011}, which act on the neutral atoms similarly to magnetic fields on charged particles. The realization of the Hofstadter model in two dimensions \cite{JakschZoller2003, AidelsburgerBloch2013,MiyakeKetterle2013,AidelsburgerGoldman2014} or on a ladder geometry \cite{AtalaBloch2014} and of the Haldane model \cite{JotsuEsslinger2014} have enabled the investigation of topological insulators in quantum gases. 

Recently, the self-organization of an artificial magnetic field in a coupled cavity-atom setup has been proposed by us using a novel coupling mechanism based on a cavity-assisted tunneling \cite{KollathBrennecke2016}. This process is induced by a Raman transition involving the dynamical cavity field and a transverse pump field. Using a running-wave pump beam, a spatially-dependent phase can be imprinted onto the atomic wave-function. In Ref.~\cite{KollathBrennecke2016} we have shown that for a phase imprint of $\varphi=\pi/2$ a self-organization of an artificial magnetic field by the feedback of the atoms and the cavity mode arises which in some limits can be described by an effective Hofstadter model. As a consequence, a chiral phase of the fermionic atoms forms. In the present work we extend the results of Ref.~\cite{KollathBrennecke2016} and map out the steady state diagram of the self-organized phases for different fillings and different magnetic fluxes. Additionally, we give a more detailed description of the solution procedure, the properties of the arising phases, and the direct detection of the chiral current via the photon losses. 

In section \ref{sec:system} we describe the combined cavity-atom setup and introduce its theoretical description. In particular, we adiabatically eliminate the cavity mode and derive an effective Hamiltonian of the fermionic atoms which needs to be solved together with a self-consistent equation in order to obtain information on the existence of a self-organized non-trivial phase with finite cavity occupation. In section \ref{sec:effHf} we discuss the properties of this effective Hamiltonian and in section \ref{sec:self} the solution of the self-consistent problem is presented. The properties of the self-organized state are discussed in section \ref{sec:chiral} focusing on the cavity occupation and the arising chiral current. The detection of the chiral current via the cavity field is described in section \ref{sec:detection}. Details of the theoretical treatment are given in the Appendix \ref{app:derivation}.
\section{Description of the setup}
\label{sec:system}
\begin{figure}
\includegraphics[width=0.7\linewidth]{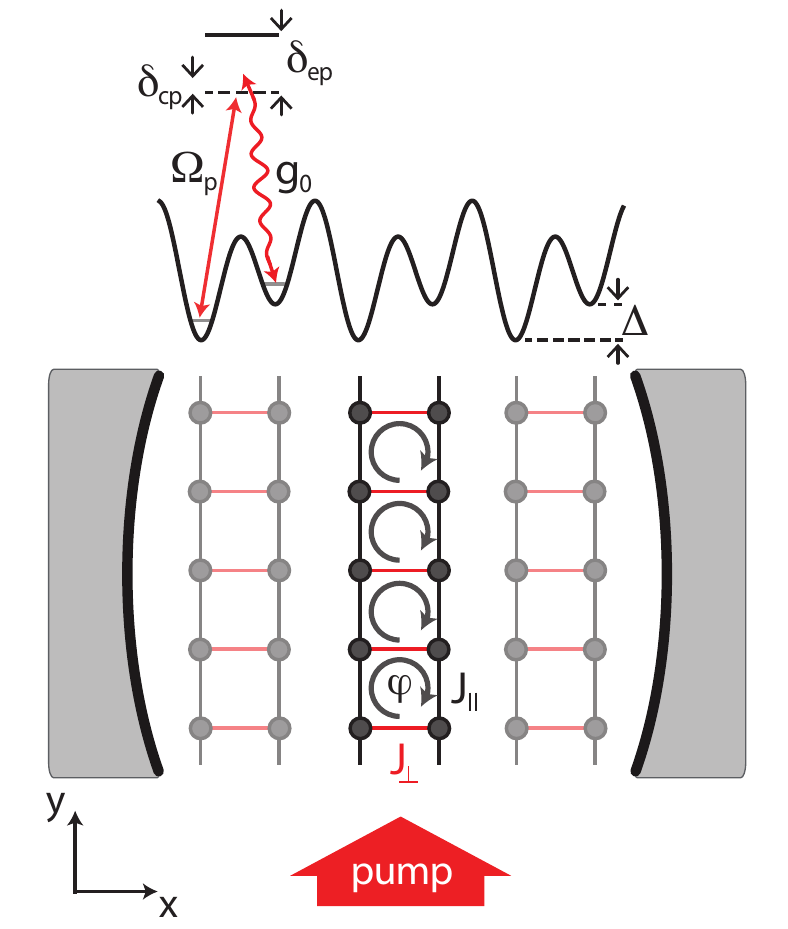}
\caption{(color online) Sketch of the setup. Fermionic atoms in an optical cavity are subjected to an optical lattice potential (not shown) which creates an array of ladders (see lower part) for which the tunneling amplitude along the legs is $J_\parallel$. The tunneling along the rungs is strongly suppressed initially by a potential offset $\Delta$ between neighboring wells. It is restored by a Raman process using a transverse pump laser beam and a cavity mode. The running-wave nature of the pump laser beam imprints a spatially dependent phase onto the atoms tunneling along the $y$-direction. This induces a dynamical artificial magnetic field with flux $\varphi$ per unit cell.}
\label{fig:cavity}
\end{figure}

We study a spin-polarized fermionic gas placed in an optical cavity and additionally subjected to optical lattice potentials (Fig.~\ref{fig:cavity}). The optical lattice potentials are chosen such that the atoms are confined to decoupled ladders. To form this structure, a strong optical lattice potential is applied along the $z$-direction to create decoupled two-dimensional layers. A second optical lattice along the $y$-direction of wavelength $\lambda_y$ induces a periodic potential with lattice spacing $d_{\parallel}=\lambda_y/2$. The lattice height along the $y$-direction is chosen sufficiently low, to allow tunneling between neighboring sites with amplitude $J_\parallel$. An additional bi-chromatic lattice potential along the $x$-direction is formed by two laser beams with wavelength $\lambda_x$ and $2\lambda_x$. The phase difference between the two laser beams is chosen such that the final lattice potential consists of an imbalanced superlattice formed of decoupled double wells with potential offset $\Delta$ as sketched in Fig.~\ref{fig:cavity}. The resulting geometry is an array of decoupled ladders where the lattice spacing between the two sites on a rung, i.e.~the double well, is denoted by $d_{\perp}$. The potential offset $\Delta$ suppresses the tunneling along the rungs.
The tunneling along the rungs can be restored using a near-resonant Raman process. The Raman process is induced by a standing-wave cavity mode with frequency $\omega_c$ and wave-vector ${\bf k}_c=k_c {\bf e}_x$ along the $x$-direction and a running-wave pump laser beam with frequency $\omega_p$ and wave-vector ${\bf k}_p=k_p {\bf e}_y+k_{p,z} {\bf e}_z$ transverse to the cavity direction. Here ${\bf e}_i$ denote the unit vectors along the direction $i=x,y,z$. The tilt of the pump laser out of the $xy$-plane can be used in order to change the in-plane component of the wave-vector $k_p$ independently of its frequency. The frequency difference $\omega_{cp}=\omega_c-\omega_p$ is chosen close to resonance with the potential offset $-\Delta/\hbar$ which induces a cavity-assisted tunneling along the rungs of the ladders. The cavity mode and the pump mode are considered to be far detuned from the internal atomic transition frequency $\omega_e$, i.e. $\omega_e\gg \omega_{c},\omega_{p}$ compared to the atomic line width. All other cavity modes are assumed to be much further detuned from possible transitions and are therefore not considered.  

The Raman transition imprints a spatially dependent phase factor $e^{-i\Delta {\bf k}\cdot{\bf r}}$ onto the atomic wave-function, where the wave-vector difference is given by $\Delta{\bf k}=\pm k_c{\bf e_x}+k_p {\bf e_y}$ and $\cdot$ denotes the scalar product. For sufficiently strong confinement along the $z$-direction, momentum transfer of the pump beam out of the $xy$-plane can be neglected. The spatially dependent phase imprint implies that if the atoms tunnel once around a plaquette of the ladder, they collect a phase $k_p d_\parallel(j+1)=\varphi(j+1)$ on the rung $j+1$ and a phase $-\varphi j$ on the rung $j$, such that the total phase is $\varphi=k_p d_\parallel=\pi \frac{\lambda_y}{\lambda_p}$, where $\lambda_p=\frac{2\pi}{k_p}$. The phases imprinted by the cavity photon do not contribute to the total phase enclosed by a plaquette.
The phase imprint on the atoms  has the same effect as a magnetic field for charged particles oriented perpendicularly to the ladder surface. Thus, in the presence a finite cavity field amplitude the atoms experience an artificial magnetic field. The value of the flux $\varphi$ depends on the projection  $k_p$ of the wave-vector of the pump laser beam onto the $y$-direction. 

In the described setup the electronically excited atomic state is almost unoccupied and can be adiabatically eliminated as described in more detail in appendix \ref{app:derivation}. Additionally, an expansion of the fermionic field in the Wannier basis of the optical lattice can be performed. This has the advantage that only the most important processes (up to neighboring lattice sites) can be considered leading to a simplified tight-binding description of the model.  
 An effective Hamiltonian can be derived which we specify for notational simplicity for one of the decoupled ladders \cite{KollathBrennecke2016}
\begin{eqnarray}
\label{eq:Heff}
H&=&H_c+H_\parallel+H_{ac} \\
H_c&=& \hbar\delta_{cp} a^\dagger a\nonumber\\
H_\parallel&=&-J_\parallel \sum_{j,m=0,1} (c_{m,j}^\dagger c_{m,j+1} + c_{m,j+1}^\dagger c_{m,j})\nonumber\\
H_{ac}&=& -\hbar\tilde{\Omega} (a K_\perp + a^\dagger K_\perp^\dagger)\nonumber\\
K_\perp&=&  \sum_{j} e^{i\varphi j}c_{0,j}^\dagger c_{1,j}. \nonumber
\end{eqnarray}
Here $a$ ($a^\dagger$) is the annihilation (creation) operator of a cavity photon in a frame rotating at frequency $\omega_p-\Delta/\hbar$ and $H_c$ is the Hamiltonian for the cavity mode in this frame with $\delta_{cp}=\omega_{cp}+\Delta/\hbar$. The operator $c_{m,j}$ ($c_{m,j}^\dagger$) is the annihilation (creation) operator of a fermion on site $j$ and leg $m=0,1$. $H_\parallel$ describes the dynamics of fermions along the legs, where $J_\parallel$ is the tunneling amplitude. $J_\parallel$ can be tuned by the intensity of the lattice potentials. 
 $H_{ac}$ encodes the dynamics along the rung induced by the Raman process via the cavity and the pump beam. The amplitude of the process is given by $\hbar\tilde{\Omega}=\frac{\hbar\Omega_p g_0}{\omega_e-\omega_p}\phi_\parallel\phi_\perp$, where $\Omega_p$ is the Rabi frequency of the pump beam and $g_0$ is the vacuum-Rabi frequency of the cavity mode. The effective parameters $\phi_\parallel$ and $\phi_{\perp}$ can be tuned via the geometry of the optical lattice and the cavity mode (see appendix \ref{app:derivation}). The operator $K_\perp$ represents the directed tunneling along the rungs of the ladders with spatially dependent phase imprint. 

Additionally to the unitary dynamics induced by the effective Hamiltonian, cavity losses lead to dissipative dynamics. The losses can be accounted for in a Lindblad master equation. The evolution of an operator $O$ can be represented by $$\frac{\partial}{\partial t} O= \frac{i}{\hbar} [H, O]+\mathcal{D}( O).$$ The dissipator is given by $\mathcal{D}( O)=\kappa\left(2 a^\dagger O a -a^\dagger a  O- O a^\dagger a\right)$ which describes the loss of cavity photons. The application of this equation to the dynamics of the expectation value of the annihilation operator of the cavity field yields
\begin{eqnarray}
i \partial_t \langle a\rangle=-\tilde{\Omega}\langle K_\perp^\dagger \rangle +( \delta_{cp}- i \kappa ) \langle a\rangle.
\end{eqnarray}
Since the time scale of the cavity field dynamics is typically fast compared that of the atomic motion, the expectation value of the cavity photon reaches rapidly a steady state and can be eliminated adiabatically. The stationary condition $\partial_t \langle a\rangle=0$ leads to the steady state value $\alpha=\langle a\rangle=\frac{\tilde{\Omega}}{\delta_{cp} - i\kappa}\langle K_\perp^\dagger \rangle$. In the experiment, the phase of the expectation value of $\aver{a}$ (U(1) symmetry) will be spontaneously broken and we will in the following consider without loss of generality the case $\aver{K_\perp}>0$. 

Using a mean-field decoupling of the atomic and cavity degrees of freedom in the equations of motion, we obtain
\begin{eqnarray}
i \hbar \partial_t \langle c_{0,j}\rangle&=& -J_\parallel \aver{c_{0,j+1} + c_{0,j-1}}-\hbar \tilde{\Omega}\aver{a} e^{i\varphi j} \aver{c_{1,j}} \nonumber\\
\end{eqnarray}
and analogous equations for $\aver{c_{1,j}}$. 
Substituting the stationary expectation value for the cavity field into the fermionic equation of motion leads to 
\begin{eqnarray}
i \hbar \partial_t \langle c_{0,j}\rangle&=& -J_\parallel \aver{c_{0,j+1} + c_{0,j-1}}-(J_\perp+iJ_I) e^{i\varphi j} \aver{c_{1,j}} \nonumber\\
 \textrm{with } J_\perp&=&\frac{\hbar\tilde{\Omega}^2\delta_{cp}}{\delta_{cp}^2 +\kappa^2}\langle K_\perp \rangle\nonumber\\
\textrm{and } J_I&=&-\frac{\hbar\tilde{\Omega}^2\kappa}{\delta_{cp}^2 +\kappa^2}\langle K_\perp \rangle.
\end{eqnarray}
In the partition of the prefactor of the last term, we have used our assumption that $\aver{K_\perp}$ is real. In the following we neglect the imaginary part of the last term, i.e.~the term proportional to $J_I$ which gives rise to dissipative dynamics. This is justified if $\kappa\ll \delta_{cp}$ and at not too long times. The resulting fermionic dynamics can be described by an effective Hamiltonian  
\begin{eqnarray}
H_F&=&H_\parallel+H_\perp \label{eq:eff_ham}
\\
H_\parallel&=&-J _\parallel \sum_{j,m=0,1} (c_{m,j}^\dagger c_{m,j+1} + c_{m,j+1}^\dagger c_{m,j})\nonumber\\
H_\perp&=& -J_\perp K_\perp + h.c. \nonumber
\end{eqnarray}
The effective hopping along the rung of the ladders needs to be determined self-consistently and is given by  
\begin{equation}
\label{eq:self}
J_\perp=A\langle K_\perp \rangle
\end{equation}
 with $A=\frac{\hbar \tilde{\Omega}^2\delta_{cp}}{\delta_{cp}^2+\kappa^2}$.

In the next section \ref{sec:effHf} we will discuss the properties of the system described by $H_F$ considering $J_\perp$ as a fixed parameter, before in section \ref{sec:self} we determine the solutions of the self-consistency equation (\ref{eq:self}). 
\section{Properties of the effective fermionic Hamiltonian}
\label{sec:effHf}
In this section we discuss the properties of the effective fermionic Hamiltonian $H_F$ (Eq.~\ref{eq:eff_ham}) considering the rung tunneling amplitude $J_\perp$ as a fixed parameter. We first introduce the Bogolioubov transformation in order to obtain the eigen-energy bands and determine the possible structures of the arising Fermi-surfaces at different fillings. Further, we determine the dependence of the expectation value of the rung tunneling $\aver{K_\perp}$ and the chiral current on the rung tunneling amplitude $J_\perp$. The former will be utilized in section \ref{sec:self} to determine the self-consistent solution. 
\subsection{Band structure and geometry of the Fermi-surfaces}  
\begin{figure*}
\includegraphics[width=0.92\textwidth]{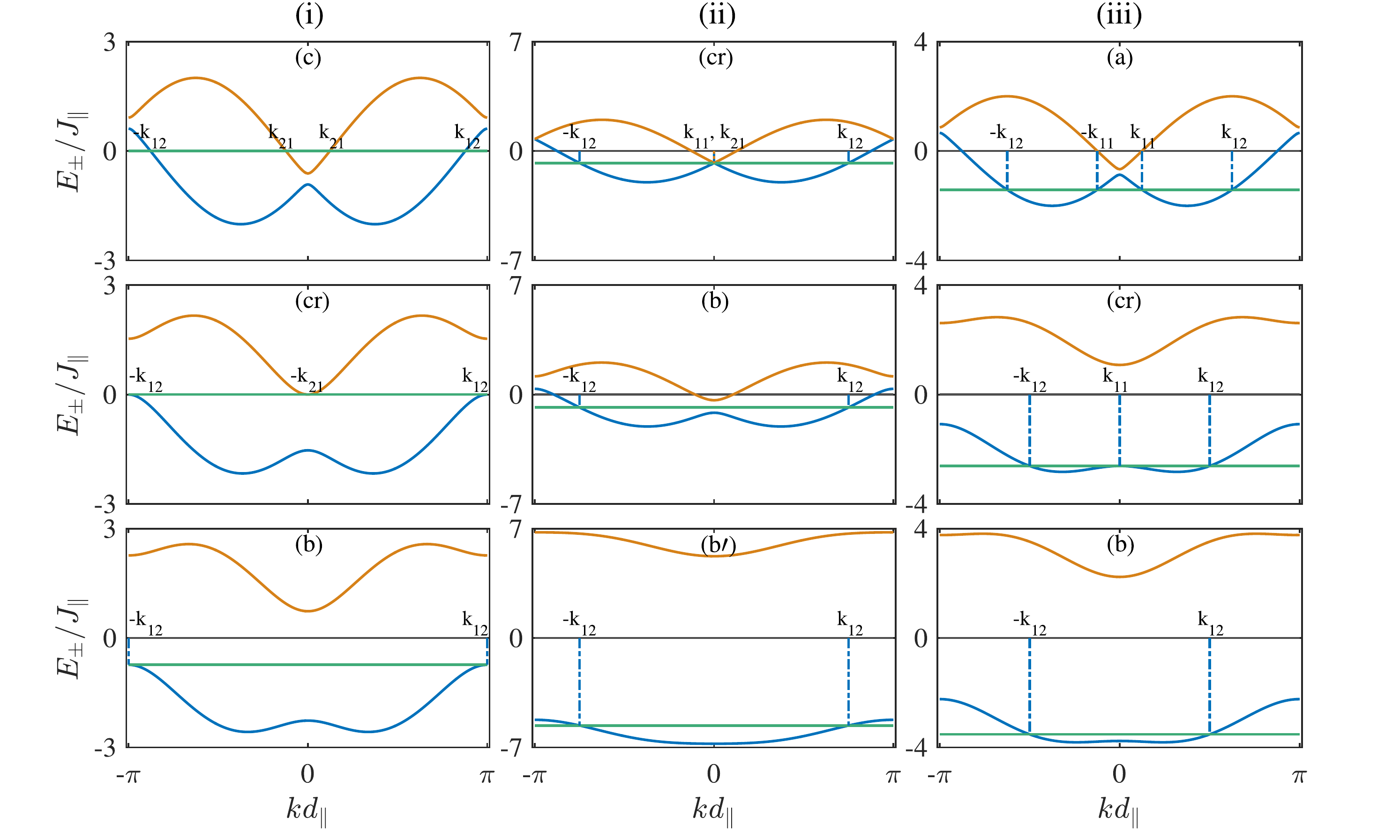}
 \caption{(Color online) 
Quasi-particle energy bands $E_-(k)$ (blue, dark line) and $E_+(k)$ (orange, light line) for $\varphi=\frac{3\pi}{4}$ and different values of the tunneling ratio $J_\perp/J_\parallel$.  The horizontal (green) line indicates the Fermi-level. The crossings with the energy bands give various Fermi points.  Column (i): $n=\frac{1}{2}$ and  $J_\perp/J_\parallel\approx 0.15, 0.77, 1.5$ from top to bottom. Column (ii): $n=\frac{3}{8}$ and $J_\perp/J_\parallel=0, 0.4, 6 $ from top to bottom.
Column (iii): $n=\frac{1}{4}$ and $J_\perp/J_\parallel\approx 0.1, 1.85, 3$ from top to bottom. The values of the tunneling ratio are chosen to exemplify the different situations (a) to (c) of the Fermi-surface described in the text.
In column (i), the upper plot exemplifies situation (c) in which both bands are partially filled with four Fermi points and the lower plot situation (b) in which only the lower band is filled with two Fermi points. The central plot shows the cross-over between these two situations.  Column (ii) shows the filling $n=\frac{3}{8}$ for which the flux $\varphi=\frac{3\pi}{4}$ is critical, i.e.~$\varphi=\varphi_{cr}$.  For a finite value of the ratio of the tunneling amplitudes only the lower band is filled and possesses two Fermi points. Column (iii) represents at low ratio of the tunneling amplitudes $J_\perp/J_\parallel=0.1$ (upper plot) situation (a), in which the lower band has two minima at $k_\pm$ and four Fermi-points. At large tunneling ratios (lower plot) the curvature of the band decreases and a crossover (central plot) to situation (b), in which two Fermi-points exist, takes place.
 }
\label{fig:band_I_II_III}
\end{figure*}
For completeness, we start with diagonalizing the Hamiltonian $H_F$ (Eq.~\ref{eq:eff_ham}) by a Bogoliubov transformation as it was previously discussed in references \cite{CarrNersesyan2006,RouxPoilblanc2007,JaefariFradkin2012,HuegelParedes2014,TokunoGeorges2014}. 
For convenience we transform the fermionic operators via a Fourier transformation along the legs of the ladder, i.e.~$c_{m,kd_\parallel}=\frac{1}{\sqrt{L}}\sum_k e^{ikd_\parallel j} c_{m,j}$, where $L$ is the number of rungs. 
The Bogoliubov transformation  which diagonalizes the effective fermionic Hamiltonian $H_F$ is given by 
\begin{eqnarray}
\gamma_{+,k}=v_k c_{0,kd_\parallel+\frac{\varphi}{2}} - u_k c_{1,kd_\parallel-\frac{\varphi}{2}} \nonumber\\
\gamma_{-,k}=u_k c_{0,kd_\parallel+\frac{\varphi}{2}} + v_k c_{1,kd_\parallel-\frac{\varphi}{2}}.
\end{eqnarray}
where $\gamma_{-,k}$ and $\gamma_{+,k}$ are the destruction operators of the quasi-particles.
 The real-valued coefficients $v_k$ and $u_k$ are determined by
\begin{eqnarray}
&v_k^2&=\frac{1}{2}\left(1+\frac{2 \sin(kd_\parallel) \sin(\frac{\varphi}{2})}{\sqrt{(J_\perp/J_\parallel)^2+ 4 \sin^2(kd_\parallel) \sin^2(\frac{\varphi}{2})}}\right)\nonumber\\
&u_k^2&=\frac{1}{2}\left(1-\frac{2 \sin(kd_\parallel) \sin(\frac{\varphi}{2})}{\sqrt{(J_\perp/J_\parallel)^2+ 4 \sin^2(kd_\parallel) \sin^2(\frac{\varphi}{2})}}\right).
\end{eqnarray}
The Hamiltonian $H_F$ can be rewritten in the diagonal form
\begin{eqnarray}
H_F=
\sum _k \left( E_+(k) \gamma_{+,k}^\dagger \gamma_{+,k} +E_-(k) \gamma_{-,k}^\dagger\gamma_{-,k}\right).
\end{eqnarray}

The quasi-particle spectrum consists of two energy bands, which are given by the expressions
\begin{eqnarray}
&E_\pm&/J_\parallel=-2 \cos(kd_\parallel) \cos\left(\frac{\varphi}{2}\right)\nonumber\\
&&\qquad\quad\pm \sqrt{(J_\perp/J_\parallel)^2+ 4 \sin^2(kd_\parallel) \sin^2\left(\frac{\varphi}{2}\right)}.
\end{eqnarray} 
The energy bands for chosen values of $J_\perp/J_\parallel$ and $\varphi$ are shown in Fig.~\ref{fig:band_I_II_III}. For a vanishing rung tunneling amplitude $J_\perp=0$, two cosine-shaped energy bands arise which are shifted  by the quasi-momentum $\pm \frac{\varphi}{2d_\parallel}$ and cross at the quasi-momentum $k=0$ and at the Brillouin zone edge. Increasing the value of the ratio of the tunneling amplitude $J_\perp/J_\parallel$ leads to a splitting of the energy band crossings into avoided crossings.  Whereas for $J_\perp/J_\parallel<2\abs{\cos(\frac{\varphi}{2})}$ the two bands still overlap in energy at different momenta, for $J_\perp/J_\parallel>2\abs{\cos(\frac{\varphi}{2})}$ the two bands are well separated by an energy gap. The lower energy band has two minima for $J_\perp/J_\parallel<2\abs{\sin(\frac{\varphi}{2})\tan(\frac{\varphi}{2})}$ [e.g.~Fig.~\ref{fig:band_I_II_III}~(i)] which are located at $$k_{\pm}=\pm \frac{1}{d_\parallel} \arccos\left(\sqrt{\left(\frac{J_\perp/J_\parallel}{2\tan(\frac{\varphi}{2})}\right)^2+\cos^2\left(\frac{\varphi}{2}\right)}\right).$$ In contrast, for $J_\perp/J_\parallel>2\abs{\sin(\frac{\varphi}{2})\tan(\frac{\varphi}{2})}$ only one minimum at $k=0$ exists [see Fig.~\ref{fig:band_I_II_III}~(ii)(b$'$)].

The various forms of the energy band structure can lead to different geometries of the Fermi-surfaces. In order to calculate zero-temperature expectation values, we need to identify these geometries. In the following we concentrate on the filling $n\leq \frac{1}{2}$ and the flux $\varphi\leq \pi$ and use the symmetries of the system afterwards to infer the expectation values for $n> \frac{1}{2}$ and $\varphi> \pi$. The filling is defined by $n=N/(2L)$, where $N$ is the number of fermions on the ladder and $L$ the number of rungs of the ladder. 
Depending on the band structure and the filling there are three typical situations for the geometry of the Fermi-surface with the number of Fermi points varying between 2 and 4. In order to describe these situations we introduce the quasi-momenta $k_{11}, k_{12}$ and $k_{21}$. The interval $k_{11}\leq |k|< k_{12}$ is given by the occupied quasi-momenta in the lower energy band.  The quasi-momentum $k_{21}$ denotes the maximal quasi-momentum up to which the upper energy band is filled. The values of $k_{11}, k_{12}$ and $k_{21}$ depend on the filling $n$, the tunneling ratio $J_\perp/J_\parallel$, and the flux $\varphi$.  With the help of these quasi-momenta, we can characterize the different Fermi-surfaces:
\begin{enumerate}[(a)]
\item In the first situation only part of the lower energy band is occupied, the upper energy band is empty and four Fermi-points arise. This situation occurs if the lower energy band has two minima at finite momenta $\abs{k_\pm}>0$ and the filling is low enough such that only $k$-values close to the band minima in the intervals $0<k_{11}< \abs{k}<k_{12}$ are populated [cf.~Fig.~\ref{fig:band_I_II_III}~(iii)(a)]. The upper energy band is empty, $k_{21}=0$. The four Fermi-points $\pm k_{11},\pm k_{12}$ lie in the lower energy band and their values can be determined by the relations $E_-(k_{11})=E_-(k_{12})$ and $2\pi n = (k_{12}-k_{11}) d_\parallel$.

\item In the second situation only part of the lower band is occupied, the upper band is empty, and two Fermi-points $\pm k_{12}$ (with $k_{12}>0$) arise. 
 This situation can occur if the lower energy band has either one minimum [e.g.~Fig.~\ref{fig:band_I_II_III}~(ii)(b$'$)] or two minima [Fig.~\ref{fig:band_I_II_III}~(i)(b), (ii)(b) or (iii)(b)]. For the case of a single minimum of the lower band, this situation arises for all fillings which do not reach the upper energy band. For the case of two minima, the filling has to be large enough such that the $k=0$ quasi-momentum in the lower band lies within the Fermi-sea ($k_{11}=0$). At the same time the upper energy band needs to be empty, i.e.~$k_{21}=0$. In both cases the Fermi-points $\pm k_{12}$ are related to the filling by $2\pi n = k_{12} d_\parallel$.

\item In the third situation both energy bands are at least partially filled and four Fermi-points, two in the lower band $\pm k_{12}$ and two in the upper band $\pm k_{21}>0$ exist. For this situation to occur, the filling must be sufficiently high such that both bands are partially filled. 
For the case $n<1/2$, four Fermi-points arise [cf.~Fig.~\ref{fig:band_I_II_III}~(i)(c)]. Two of them lie in the lower energy band at $\pm k_{12}$ (with $k_{12}>0$ and $k_{11}=0$) and two in the upper energy band, $\pm k_{21}$ with $k_{21}>0$. The Fermi-points can be determined from the relations $E_+(k_{21})=E_-(k_{12})$ and $2\pi n = (k_{12}+k_{21}) d_\parallel$.
\end{enumerate}

These three typical structures are separated by 'critical' geometries. We denote the separating values of the tunneling ratio by $(J_\perp/J_\parallel)_{cr}$. The first critical geometry separates the situation (b) and (c). In this geometry the Fermi surface touches the upper band and $k_{21}$ vanishes [see Fig.~\ref{fig:band_I_II_III}~(i)(cr)]. The second critical geometry separates case (a) and (b). At this value $k_{11}$ becomes zero and the transition between four Fermi points to two Fermi points in the lower band takes place [see Fig.~\ref{fig:band_I_II_III}~(iii)(cr)]. As shown in Fig.~\ref{fig:band_I_II_III}~(ii)(cr) the two critical geometries can fall together in the particular situation that $(J_\perp/J_\parallel)_{cr}=0$, since then the lower and upper band cross. This occurs for a specific value of the flux which we denote as the critical value $\varphi_{cr}$  and which is related to the filling by $\varphi_{cr} =2\pi n $.

In the following subsections we discuss how the structure of the Fermi-surface influences physical properties such as the expectation value of the rung tunneling and the chiral current. 

\subsection{Expectation value of the rung tunneling} 
\begin{figure*}
\centering
\includegraphics[width=0.82\textwidth]{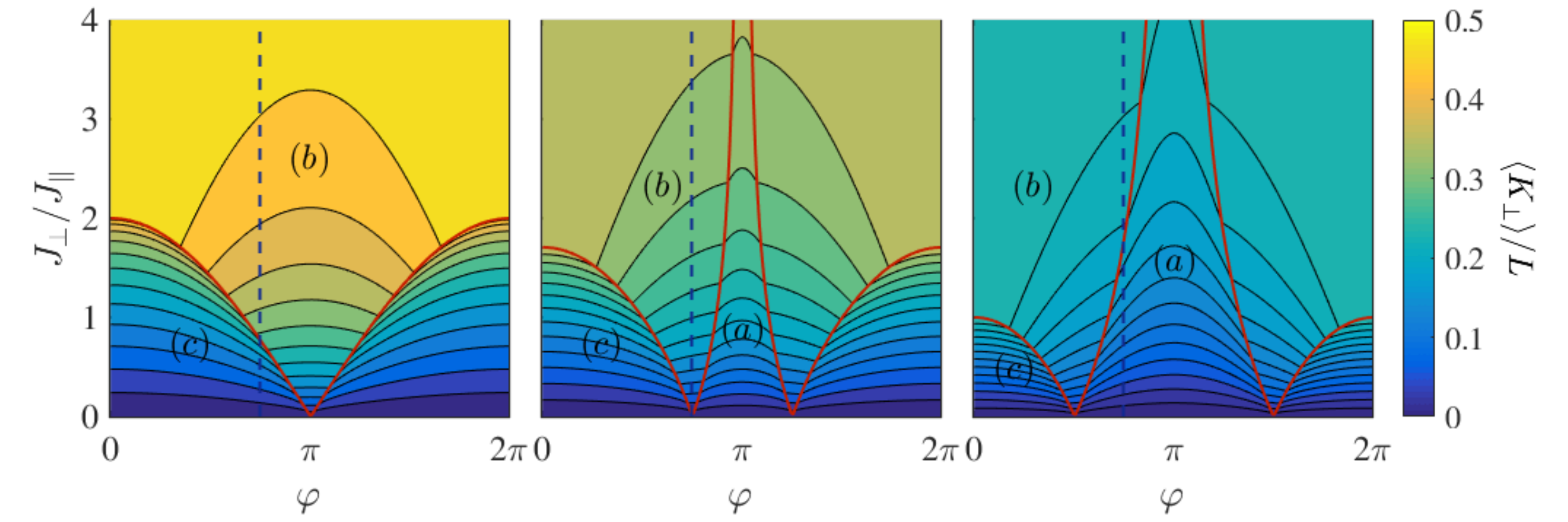}
\caption{(Color online) Dependence of the expectation value of the rung tunneling $\aver{K_\perp}/L$ on the flux $\varphi$ and the tunneling ratios $J_\perp/J_\parallel$ at fillings  $n=\frac{1}{2}$, $\frac{3}{8}$ and $\frac{1}{4}$ from left to right. The corresponding critical values of the flux are $\varphi_{cr}=\pi$, $\frac{3\pi}{4}$ and $\frac{\pi}{2}$ (and $ 2\pi-\varphi_{cr}$). The red (solid) curves show the critical values of the tunneling ratio $(J_\perp/J_\parallel)_{cr}$ as a function of flux. The dashed vertical lines mark the value of flux $\frac{3\pi}{4}$ which corresponds to Fermi-surfaces shown in Fig.~\ref{fig:band_I_II_III}. The letters mark the regions with different Fermi-surface geometries described in the text.}
\label{fig:Kperp_J_phi_n0p25}
\end{figure*}
The expectation value of the rung tunneling $\aver{K_\perp}$ in the ground state has two contributions with opposite sign for the quasi-particles in the lower and upper energy band. This can be seen in the expression 
\begin{eqnarray}
\label{eq:expKerp}
\frac{\langle K_\perp\rangle}{L} &=&\frac{1}{L} \sum_{k} u_k v_k \langle\gamma_{-,k}^\dagger \gamma_{-,k} -\gamma_{+,k}^\dagger \gamma_{+,k}\rangle
\end{eqnarray}

Thus, the discussed geometries of the Fermi-surface will have an influence on the behavior of this expectation value. 
In order to evaluate the expectation value $\langle K_\perp\rangle$ we take the continuum limit and rewrite the arising integrals as elliptic integrals  ${\mathcal F}$ of the first kind:
\begin{widetext}
\begin{eqnarray}
\langle K_\perp\rangle/L &\approx&\frac{1}{2\pi}\int_{k_{11}}^{k_{12}} \frac{(J_\perp/J_\parallel)d_\parallel}{\sqrt{(J_\perp/J_\parallel)^2+ 4 \sin^2(kd_\parallel) \sin^2(\frac{\varphi}{2})}}{\textrm d}k - \frac{1}{2\pi}\int_{0}^{k_{21}} \frac{(J_\perp/J_\parallel)d_\parallel}{\sqrt{(J_\perp/J_\parallel)^2+ 4 \sin^2(kd_\parallel) \sin^2(\frac{\varphi}{2})}} {\textrm d} k\nonumber\\
&=& \frac{1}{2\pi} \left[{\mathcal F}\left(k_{12}d_\parallel, -\frac{1}{{\tilde J}^2}\right)-{\mathcal F}\left(k_{11}d_\parallel, -\frac{1}{{\tilde J}^2}\right)-{\mathcal F}\left(k_{21}d_\parallel, -\frac{1}{{\tilde J}^2}\right) \right],
\label{eq:Kperp_cont}
\end{eqnarray}
\end{widetext}
where we defined ${\tilde J}:= \frac{J_\perp/J_\parallel}{2\sin(\frac{\varphi}{2})}$.

As a typical example of the arising behavior we show in Fig.~\ref{fig:Kperp_J_phi_n0p25} the expectation value of the rung tunneling $\langle K_\perp \rangle/L$ versus the ratio of the tunneling amplitudes $J_\perp/J_\parallel$ and flux $\varphi$ for three different fillings.  Different geometries of the Fermi-surfaces (a)-(c) are separated by the red lines. We focus first on the filling $n=1/4$ (right panel). 
The expectation value of the rung tunneling increases monotonically with increasing tunneling ratio $J_\perp/J_\parallel$. In region (c) the rung tunneling shows a very steep rise. There are four Fermi-points, two of which are situated in each energy band. By increasing the rung tunneling amplitude $J_\perp$, the upper energy band rises and the upper Fermi-points $k_{21}$ move towards the band minimum $k=0$. Thus, the contribution of the second band decreases which leads due to the negative sign in Eq.~\ref{eq:Kperp_cont} to an increase of the expectation value of the rung tunneling. A second contribution stems from the broader Fermi-surface in the lower band. The rise of the rung tunneling becomes much more moderate for large values of the tunneling ratio $(J_\perp/J_\parallel)$ in region (a) and (b), since here only the lower energy band contributes. In region (a) and (b) there are four and two Fermi-points, respectively, which are situated in lower energy band. By increasing the rung tunneling amplitude $J_\perp$ the lower energy band flattens and the resulting contributions of the filled quasi-momenta increase.  The cross-over between two regions [between (a) and (b) or between (c) and (b)] at $(J_\perp/J_\parallel)_{cr}$ shows up in a cusp. At the critical flux $\varphi_{cr}$, only situation (b) occurs. 

The shown behaviour of the expectation value of the rung tunneling for $n=1/4$ is very typical. Changing the filling mostly influences the extensions of the discussed regions. 
In particular, for increasing filling $1/4<n<1/2$ the region (a) in the center with four Fermi-points in the lower energy band shrinks until at $n=1/2$ no such region persists and $\varphi_{cr}=\pi$.
 
The symmetries of the system give the relation $\langle K_\perp(n ,2\pi-\varphi, J_\perp/J_\parallel) \rangle = \langle K_\perp(n, \varphi, J_\perp/J_\parallel) \rangle$ (with $\varphi \in [0,\pi]$) between low and high flux.  
A similar expression relating fillings higher than half filling to fillings lower than half filling can be derived. The relation is given by $ \langle K_\perp(1-n ,\varphi, J_\perp/J_\parallel) \rangle = \langle K_\perp(n, \varphi, J_\perp/J_\parallel) \rangle$ with $n<1/2$. These symmetry arguments enable us to deduce the full behavior of the expectation value of the rung tunneling from the discussed situations.
\subsection{Properties of the chiral current}
\begin{figure*}
\centering
\includegraphics[width=0.82\textwidth]{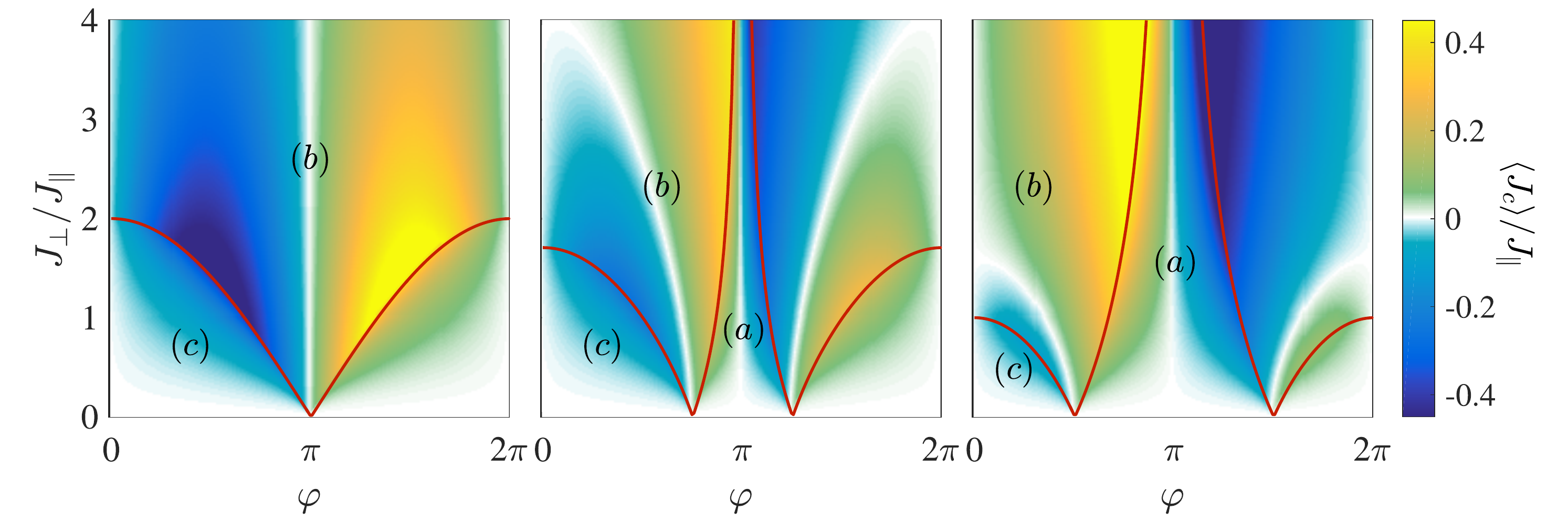}
\caption{(Color online) Dependence of the chiral current on the flux $\varphi$ and the tunneling ratios $J_\perp/J_\parallel$ at fillings  $n=\frac{1}{2}$, $\frac{3}{8}$ and $\frac{1}{4}$ from left to right. The corresponding critical values of the flux are $\varphi_{cr}=\pi$ , $\frac{3\pi}{4}$ and $\frac{\pi}{2}$ (and $ 2\pi-\varphi_{cr}$). The red curves show the critical value of tunneling ratio $(J_\perp/J_\parallel)_{cr}$ as a function of flux. The letters mark the regions with different Fermi-surface geometries described in the text.}
\label{fig:Jc_J_phi_n0p25}
\end{figure*}
\begin{figure}
\includegraphics[width=1\columnwidth,clip=true]{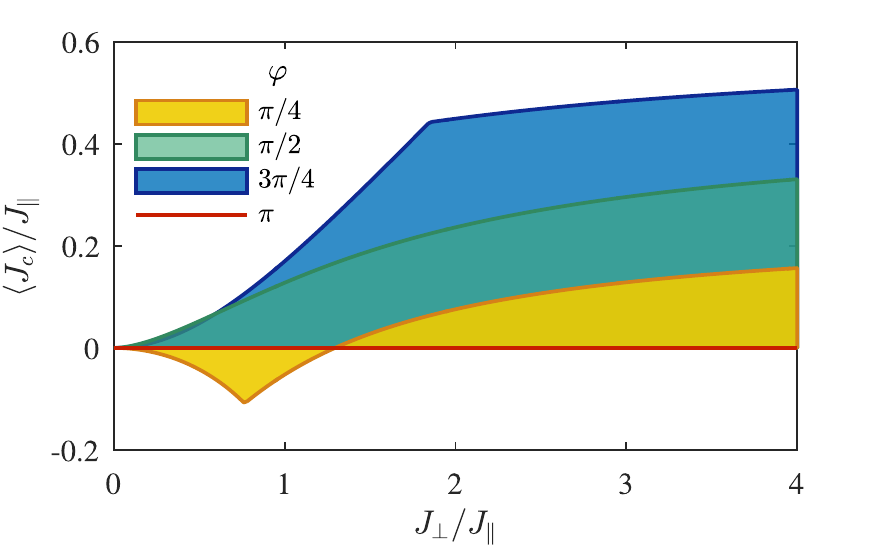}
\caption{(Color online)  Cuts through the right panel of Fig.~\ref{fig:Jc_J_phi_n0p25} showing the dependence of the chiral current $J_c$ on the tunneling ratio $J_\perp/J_\parallel$ for various values of the flux $\varphi$ at quarter filling $n=\frac{1}{4}$. 
}
\label{fig:Jc_J_phi_n0p25_cuts}
\end{figure}
One of the interesting physical effects of an artificial gauge field is the possible creation of chiral currents. On a ladder structure the chiral current is defined as the difference of the current along the two legs:
\begin{eqnarray}
J_c=\frac{1}{L-1}\sum_j (j_{0,j} - j_{1,j}). 
\end{eqnarray}
Here we used the definition of the current on leg $m$ given by 
\begin{eqnarray}
j_{m,j}= -i J_\parallel (c_{m,j}^\dagger c_{m,j+1} - c_{m,j+1}^\dagger c_{m,j}).
\end{eqnarray}
Similar to the expression for the rung tunneling, the chiral current has contributions both from the quasi-particles in the lower and upper energy band as given by the following expression
\begin{eqnarray}
&&\langle J_c\rangle/J_\parallel  =\frac{2}{L-1}\sum_k \nonumber\\ 
&&\left  [\left[\sin\left(kd_\parallel+\frac{\varphi}{2} \right)u_k^2- \sin\left(kd_\parallel-\frac{\varphi}{2}\right)v_k^2  \right]\langle\gamma_{-,k}^\dagger \gamma_{-,k}\rangle\right.\nonumber\\ 
&&+\left. \left[\sin\left(kd_\parallel+\frac{\varphi}{2} \right)v_k^2- \sin\left(kd_\parallel-\frac{\varphi}{2}\right)u_k^2  \right]\langle\gamma_{+,k}^\dagger \gamma_{+,k}\rangle\right]\nonumber\\
\end{eqnarray}
In the continuum limit the expression becomes 
\begin{widetext}
\begin{eqnarray}
&&\langle J_c\rangle/J_\parallel  \approx \frac{2}{\pi} \sin\left (\frac{\varphi}{2}\right) \left[\sin(k_{12}d_\parallel)-\sin(k_{11}d_\parallel)+\sin(k_{21}d_\parallel) \right]\nonumber \\
&+&\frac{4}{\pi}{\tilde J}\cos\left(\frac{\varphi}{2}\right) \left[{\mathcal E}\left(k_{11}d_\parallel, -\frac{1}{{\tilde J}^2}\right)-{\mathcal F}\left(k_{11}d_\parallel, -\frac{1}{{\tilde J}^2}\right)+{\mathcal E}\left(k_{21}d_\parallel, -\frac{1}{{\tilde J}^2}\right)-{\mathcal F}\left(k_{21}d_\parallel, -\frac{1}{{\tilde J}^2}\right)\right. \nonumber \\ 
&&\left. -{\mathcal E}\left(k_{12}d_\parallel, -\frac{1}{{\tilde J}^2}\right)+{\mathcal F}\left(k_{12}d_\parallel, -\frac{1}{{\tilde J}^2}\right)\right],
\end{eqnarray}
\end{widetext}
where ${\mathcal E}$ denotes the elliptic integral of second kind. 

To gain insight into the typical behavior of the chiral current, Fig.~\ref{fig:Jc_J_phi_n0p25} shows the chiral current $\langle J_c\rangle /J_\parallel$ versus the ratio of the tunneling amplitudes $J_\perp/J_\parallel$ and flux $\varphi$ at different fillings. As in Fig.~\ref{fig:Kperp_J_phi_n0p25} we have added the lines separating the different geometries of the Fermi-surfaces (a)-(c). 
Additionally, Fig.~\ref{fig:Jc_J_phi_n0p25_cuts} presents cuts at various values of the flux for $n=1/4$. 
For $n=1/4$ and $\varphi<\varphi_{cr}=\pi/2$ within region (c) the chiral current takes a decreasing negative value with increasing value of the ratio of the tunneling amplitudes until it reaches the boundary to region (b). At the boundary between the two regions the chiral current reaches its maximally negative value. In region (b), with increasing ratio of the tunneling amplitudes $J_\perp/J_\parallel$, the chiral current increases and even changes its sign which means that it inverts its direction. For larger values of the flux inside region (a)  the chiral current shows a steep rise for intermediate values of the tunneling ratio. This increase crosses over with a cusp at the boundary to region (b) in which the rise becomes much more moderate. At critical flux $\varphi_{cr}$ which always lies in region (b) the chiral current shows a monotonic increase with increasing tunneling ratio. The described behavior is again typical and it is sustained for different fillings, where mainly the location of the different regions changes.  

For $\varphi>\pi$ the same behavior occurs in inverse order with negative sign. This is due to the symmetry of the system which leads to the relation $\langle J_c(n ,2\pi- \varphi, J_\perp, J_\parallel) \rangle = -\langle J_c(n, \varphi, J_\perp, J_\parallel) \rangle$.
Fillings above half filling can be inferred from the relation $\langle J_c(1-n ,\varphi, J_\perp, J_\parallel) \rangle = \langle J_c(n, \varphi, J_\perp, J_\parallel) \rangle$.
\section{Self-consistent solution of the effective fermionic model}
\label{sec:self}
\begin{figure*}\centering
\includegraphics[width=1\textwidth]{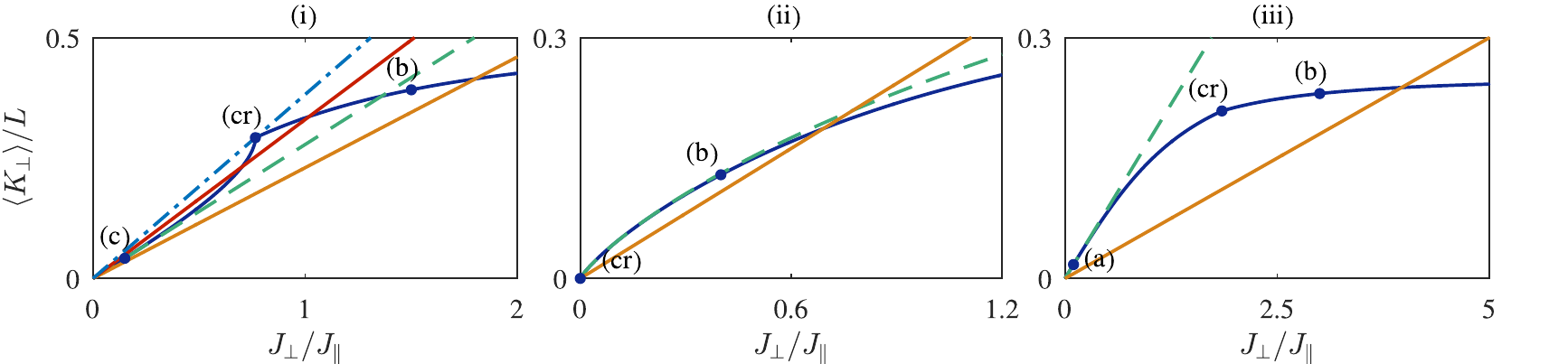}
\caption{(Color online) Graphical interpretation of the self-consistency condition at $\varphi=\frac{3\pi}{4}$ for different fillings $n=\frac{1}{2}$ corresponding to scenario (i), $n=\frac{3}{8}$ corresponding to scenario (ii), and $n=\frac{1}{4}$ corresponding to scenario (iii) described in the text. The small letters label the regions of the different geometries of the Fermi-surfaces discussed earlier and the band structure corresponding to the parameters marked by filled circles is depicted in Fig.~\ref{fig:band_I_II_III}. The blue solid line shows the left-hand side (LHS) of the self-consistency condition, i.e.~the expectation value of the rung tunneling $\langle K_\perp\rangle/L$ (cuts of dashed lines in Fig.~\ref{fig:Kperp_J_phi_n0p25}). The linear curves show the right-hand side (RHS) $\frac{J_\parallel}{AL}J_\perp/J_\parallel$ of the self-consistency condition Eq.~\ref{eq:self} for chosen values of the pump strength $A$. The crossings between the RHS and the LHS give the solutions. In panel (i) the dotted-dashed line corresponds to the minimum value $A_{cr,\text{i}}$ (see Eq.~\ref{eq:AminI}) for which a self-consistent solution exists. For intermediate values of $A$ (red curve) two solutions exist, before above the value $A_{\text {max,i}}$ (dashed line) only one non-trivial solution (orange solid line) exists. In panel (ii) there exists for each value of $A$ one non-trivial self-consistent solution as exemplified for the orange solid line. The dashed line represents the approximation Eq.~(\ref{eq:Kperp_J_II}) of the expectation value of the rung tunneling $\langle K_\perp\rangle/L$ for small ratios of the tunneling amplitudes $J_\perp/J_\parallel$.
In panel (iii), the dashed line corresponds to the minimal value $A_{cr,\text{iii}}$ (Eq.~\ref{eq:Amin_III}) for which a self-consistent solution exists and the orange solid line shows a solution which lies in region (b).
}
\label{fig:kperp_J_123}
\end{figure*}
\begin{figure}
\includegraphics[width=1\columnwidth,clip=true]{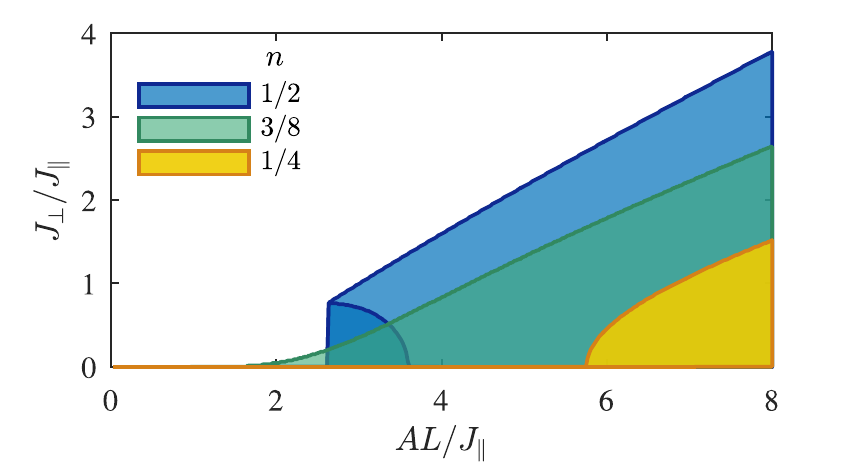}
 \caption{(Color online) The solutions for $J_\perp/J_\parallel$ of the self-consistency equation versus the pump strength $A$ for the parameters shown in Fig.~\ref{fig:kperp_J_123}.} 
\label{fig:Jperp_vs_A}
\end{figure}
After having discussed in section \ref{sec:effHf} the properties of the effective fermionic Hamiltonian $H_F$ (Eq.~\ref{eq:eff_ham}) for fixed rung tunneling amplitudes, we now turn to the solution of the self-consistent problem which includes the feedback of the cavity mode allowing $J_\perp$ to self-adjust. The gained insight into the behavior of the expectation value of the rung tunneling  $\langle K_\perp\rangle$ on $J_\perp$ will help to determine the possible solutions.

\subsection{Graphical interpretation of the self-consistency condition}
The self-consistency condition (Eq.~\ref{eq:self}) can be reformulated into the condition $\langle K_\perp(J_\perp/J_\parallel)\rangle/L=\frac{J_\parallel}{A L} J_\perp/J_\parallel$. The left-hand side of the condition contains the complicated dependence of the expectation value $\aver{K_\perp}$ on the ratio of the rung tunneling amplitudes $J_\perp/J_\parallel$, whereas the right-hand side represents a linear function of $J_\perp/J_\parallel$ with slope  $\frac{J_\parallel}{A L}$. The slope can be tuned e.g. via the pump strength $A$. This form of the condition suggests a simple graphical interpretation. Plotting both sides of the condition, the solutions are determined by the crossings of the two curves. 

Fig.~\ref{fig:kperp_J_123} shows the expectation value of the rung tunneling $\langle K_\perp\rangle/L$ at $\varphi=3\pi/4$ for three characteristic scenarios, which are marked in Fig.~\ref{fig:Kperp_J_phi_n0p25} by dashed lines.  
These correspond to (i) the crossing between region (c) and (b) with increasing tunneling ratio, (ii) the special situation that one remains within region (b) for all values of the tunneling ratio (i.e.~the flux corresponds to the critical flux $\varphi_{cr}$), and (iii) the crossing between region (a) and (b) for increasing tunneling ratio. We note that the scenarios are very typical and could also be realized at a fixed filling by varying the flux. 

\begin{itemize}
\item[(i)] In Fig.~\ref{fig:kperp_J_123}~(i) the filling is chosen such that at small values of the ratio of the tunneling $J_\perp/J_\parallel$ the system has the Fermi surface structure (c), i.e.~two Fermi points in the lower and two in the upper energy band, whereas at larger values it crosses over to situation (b) with two Fermi points in the lower energy band [cf.~Fig.~\ref{fig:band_I_II_III}~(i)]. The resulting expectation value of the rung tunneling $\langle K_\perp\rangle/L$ has a concave curvature in region (c) below $(J_\perp/J_\parallel)_{cr}$ and a convex curvature in region (b) above $(J_\perp/J_\parallel)_{cr}$ with a cusp at the critical tunneling. Thus, no solution exists below a critical pump strength $A_{cr,\text{i}}$ which relates to the critical value of the pump strength (see dotted-dashed line in Fig.~\ref{fig:kperp_J_123}~(i)).
\begin{equation}
\label{eq:AminI}
A_{cr,\text{i}}\frac{ L}{J_\parallel}=\frac{\langle K_\perp\rangle_{cr}/L}{(J_\perp/J_\parallel)_{cr}},
\end{equation}
 where $\langle K_\perp\rangle_{cr}/L$ is the value of $\langle K_\perp\rangle/L$ evaluated at the critical hopping $\left(\frac{J_\perp}{J_\parallel}\right)_{cr}$. For the shown parameters in Fig.~\ref{fig:Jperp_vs_A} at $n=1/2$, the critical value of the pump strength is $A_{cr,i} \approx 2.63 J_\parallel/L $.
Over a certain regime of values of $A>A_{cr,\text{i}}$ two solutions exist which signals a possible bistability. The first solution is always above the critical value of the tunneling ratio $(J_\perp/J_\parallel)_{cr}$ (situation (b)). The solution $J_\perp$ grows monotonically and persists even for large values of the parameter $A$. At large values of $A$, it can be approximated by a linear growth $J_\perp \approx AL d_\parallel k_F/(2 \pi)$. 

In contrast, the second solution decreases with increasing $A$ (in situation (c)) and only exists up to a value $\left(A \right)_{\text{max,i}}$  which is related to the slope of the expectation value of the  rung tunneling at small  $(J_\perp/J_\parallel)$ [cf.~dashed line in Fig.~\ref{fig:kperp_J_123}~(i)]. By expanding Eq.~\ref{eq:Kperp_cont} for $J_\perp/J_\parallel\ll 1$ this upper limit of pump strength is calculated for $0\leq\varphi\leq\pi$ to be given by 
\begin{eqnarray}
A_{\text{max,i}}\frac{ L}{J_\parallel}=\frac{4\pi \sin(\frac{\varphi}{2})}{\log\left(\frac{\tan(\frac{n\pi}{2}+\frac{\varphi}{4})}{\tan(\frac{n\pi}{2}-\frac{\varphi}{4})}\right)}.
\label{eq:Amax_I2}
\end{eqnarray}
For the shown parameter in Fig.~\ref{fig:Jperp_vs_A} at $n=1/2$, the maximal value of the pump strength is $A_{\text{max,i}}\approx 3.59 J_\parallel/L$.
In the regime of coexistence of the two solutions, a stability analysis of the different solutions has to be performed in order to decide which of these is taken beyond adiabatic elimination. Exact numerical calculations for small system sizes point towards the stability of the first solution, i.e.~the solution above the critical value of the tunneling $(J_\perp/J_\parallel)_{cr}$ and an instability of the second solution \cite{KollathBrennecke2016}. 

Let us emphasize, that the described scenario (i) applies to all parameter sets in which a direct crossing between region (c) and (b) takes place. Mostly, the values of the solution and the location of the critical and maximal value of $A$ changes. Since the effective tunneling ratio of the self-consistent solution is proportional to the mean cavity field amplitude $\alpha$, the results show that a sudden occupation  of the cavity field takes place at the critical pump strength $A_{cr,i}$. This indicates the self-organization of a non-trivial symmetry-broken state. The properties of this state and in particular of the atomic gas will be investigated in more detail in the next subsection.

\item[(ii)] The second scenario is shown in Fig.~\ref{fig:kperp_J_123}~(ii), where the filling is chosen such that $\varphi=3\pi/4$ corresponds to the critical flux $\varphi_{cr}$. 
In this case all values of $J_\perp/J_\parallel>0$ lie above $(J_\perp/J_\parallel)_{cr}=0$ and situation (b) is realized, i.e.~two Fermi points $\pm k_{12}>0$ exist in the lower band [cf.~Fig.~\ref{fig:band_I_II_III}~(ii)]. The expectation value of the rung tunneling $\langle K_\perp\rangle/L$ has a convex curvature. To be more precise, the expectation value of the rung tunneling for small tunneling ratio at a fixed finite filling can be expanded as 
\begin{eqnarray}
\left(\langle K_\perp\rangle/L\right)_{\text{ii}}=&&\frac{ \log\left(\frac{8 J_\parallel \sin^2(n\pi)}{J_\perp\cos(n\pi)}  \right)}{4\pi \sin(n\pi)}J_\perp/J_\parallel \nonumber\\
&&+O\left[\left(J_\perp/J_\parallel\right)^3\right],
\label{eq:Kperp_J_II}
\end{eqnarray}
which shows a logarithmic convex behavior for small tunneling ratios [dashed line in Fig.~\ref{fig:kperp_J_123}~(ii)]. Thus, since the derivative at low values of the tunneling ratio diverges, for all finite values of the pump strength $A>0$ a self-consistent solution arises as seen in Fig.~\ref{fig:Jperp_vs_A} at $n=3/8$. Due to the overall convex form of the expectation value of the rung tunneling for each value of $A$ a single self-consistent solution exists. Typically for small $A$, the solution $J_\perp$ increases slowly with increasing the pump strength $A$. This leads also to a slow increase of the cavity field amplitude with the applied transverse pump strength which has to be contrasted with the sudden onset in scenario (i). 

The scenario (ii) is much more rare than the previously discussed scenario (i), since it only exists at the critical flux of a chosen filling. 

\item[(iii)] The third scenario is shown in Fig.~\ref{fig:kperp_J_123}~(iii), where the filling is chosen, such that the flux fulfills $\varphi_{cr}< \varphi=3\pi/4<\pi$. This means that at small ratios of the tunneling amplitudes the system is in situation (a) and crosses over to situation (b) at larger ratios.
The form of the expectation value of the rung tunneling $\aver{K_\perp}$ increases with increasing tunneling ratio and has a slight cusp at the critical value between region (a) and (b). Since the curve grows at large value of the tunneling ratio monotonically, there exists no upper limit for the value of $A$ for which a non-trivial self-consistent solution arises. However, the low values of the tunneling ratio need to be considered more carefully. In particular, the expansion of the expectation value of the rung tunneling in this limit behaves linear followed by a bending down at larger values of $J_\perp/J_\parallel$. Thus, below a lower critical value $A$ no solution exists. This critical value $\left(A \right)_{cr,\text{iii}}$ is given by 
\begin{eqnarray}
\label{eq:Amin_III}
\left(A\right)_{cr,\text{iii}}=\frac{4\pi \sin(\frac{\varphi}{2})J_\parallel}{L \log\left(\frac{\tan(\frac{\varphi}{4}+\frac{n\pi}{2})}{\tan(\frac{\varphi}{4}-\frac{n\pi}{2})}\right)},
\end{eqnarray}
where we have considered $0\leq\varphi\leq\pi$ [e.g.~see dashed line in Fig.~\ref{fig:kperp_J_123}~(iii)]. This means that a single self-consistent solution exists for all values of $A>A_{cr,\text{iii}}$ as seen for $n=1/4$ in Fig.~\ref{fig:Jperp_vs_A}. In scenario (iii) the solution for $J_\perp$ slowly increases with $A$ and no sudden jump of the cavity field amplitude is found. 
\end{itemize}

To summarize, for scenario (i) and (iii) a lower critical value $A_{cr}$ exists below which only the trivial solution of an empty cavity exists. Above this critical value, at least one non-trivial solution arises. In contrast, at the critical flux, scenario (ii) occurs for which a single solution arises for all finite values of the pump strength $A$.
\subsection{Self-organized chiral state}
\label{sec:chiral}
\begin{figure}
\centering
\includegraphics[width=0.5\textwidth]{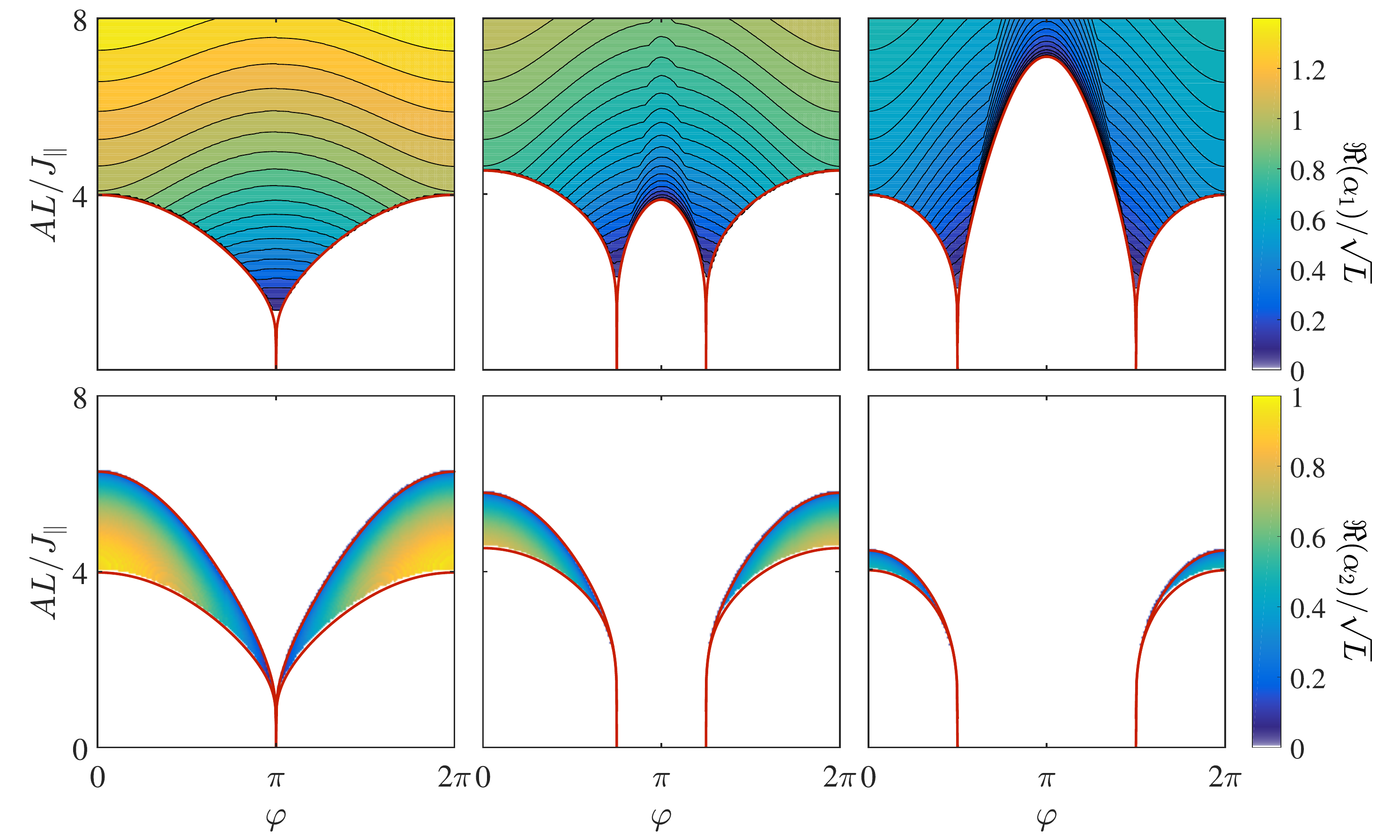}
\caption{(Color online)  Regime of existence of a self-organized chiral state for fillings $n=\frac{1}{2},\frac{3}{8},\frac{1}{4}$ from left to right. Upper panels: The red solid curves mark the critical values of pump strength $A_{cr}$ above which a self-consistent solution with a finite cavity field exists. Lower panels: the regime of existence of the second solution of scenario (iii) (lying in region c) is represented in between the lower and upper red solid line. In the self-organized phase the real part of the expectation value $\Re(\alpha)/\sqrt{L}$ of the cavity field is shown by the color code. Here, $\hbar \delta_{cp}= J_\parallel$ and $\hbar\kappa=0.05 J_\parallel$. The subscripts denote the two solutions of the self-consistency equation.}
\label{fig:alpha_phi_A_diffn}
\end{figure}
\begin{figure}
\centering
\includegraphics[width=0.5\textwidth]{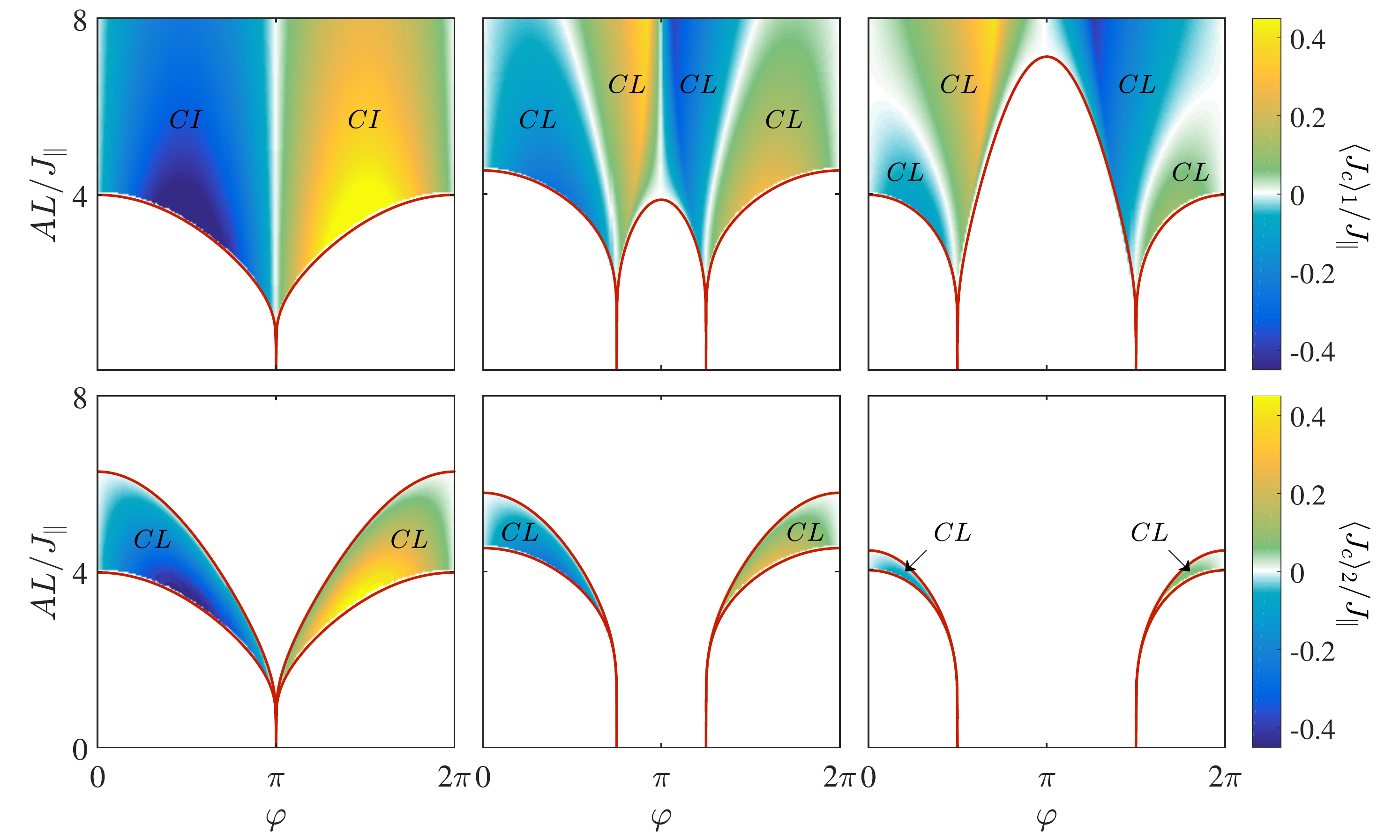}
\caption{(Color online)  Regime of existence of a self-organized chiral state for fillings $n=\frac{1}{2},\frac{3}{8},\frac{1}{4}$ from left to right. Upper panels: The  red solid curves mark the critical values of pump strength $A_{cr}$ above which a self-consistent solution with a finite cavity field exists. Lower panels: the regime of existence of the second solution of scenario (iii) (lying in region c) is represented in between the lower and upper red solid line. In the self-organized phase the expectation value of the chiral current is shown by the color code for the respective self-consistent solution. A chiral insulator (liquid) is denoted by CI (CL), respectively. The lines where the chiral current  vanishes (see color coding) correspond to a normal insulator and liquid according to the bordering phases. Parameters as in Fig.~\ref{fig:alpha_phi_A_diffn}. The subscripts denote the two solutions of the self-consistency equation.
}
\label{fig:Jc1_phi_A_diffn}
\end{figure}
\begin{figure}
\centering
\includegraphics[width=0.5\textwidth]{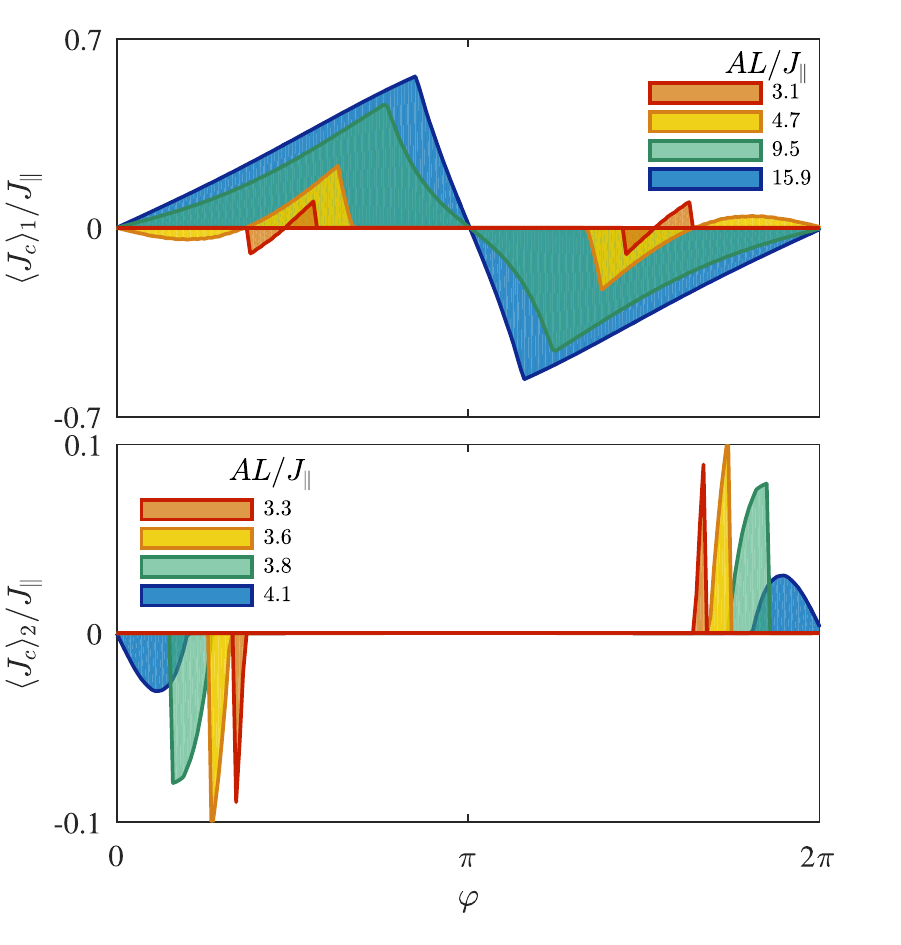}
\caption{(Color online) Chiral current versus flux for different pump strength $A$. Shown are different cuts through the density plot  Fig.~\ref{fig:Jc1_phi_A_diffn} at quarter filling $n=\frac{1}{4}$ for the first (top) and second (bottom) self-consistent solution. Subscripts label the two solutions of the self-consistency equation.}
\label{fig:Jc12_phi_A}
\end{figure}
\begin{figure}
\includegraphics[width=0.5\textwidth]{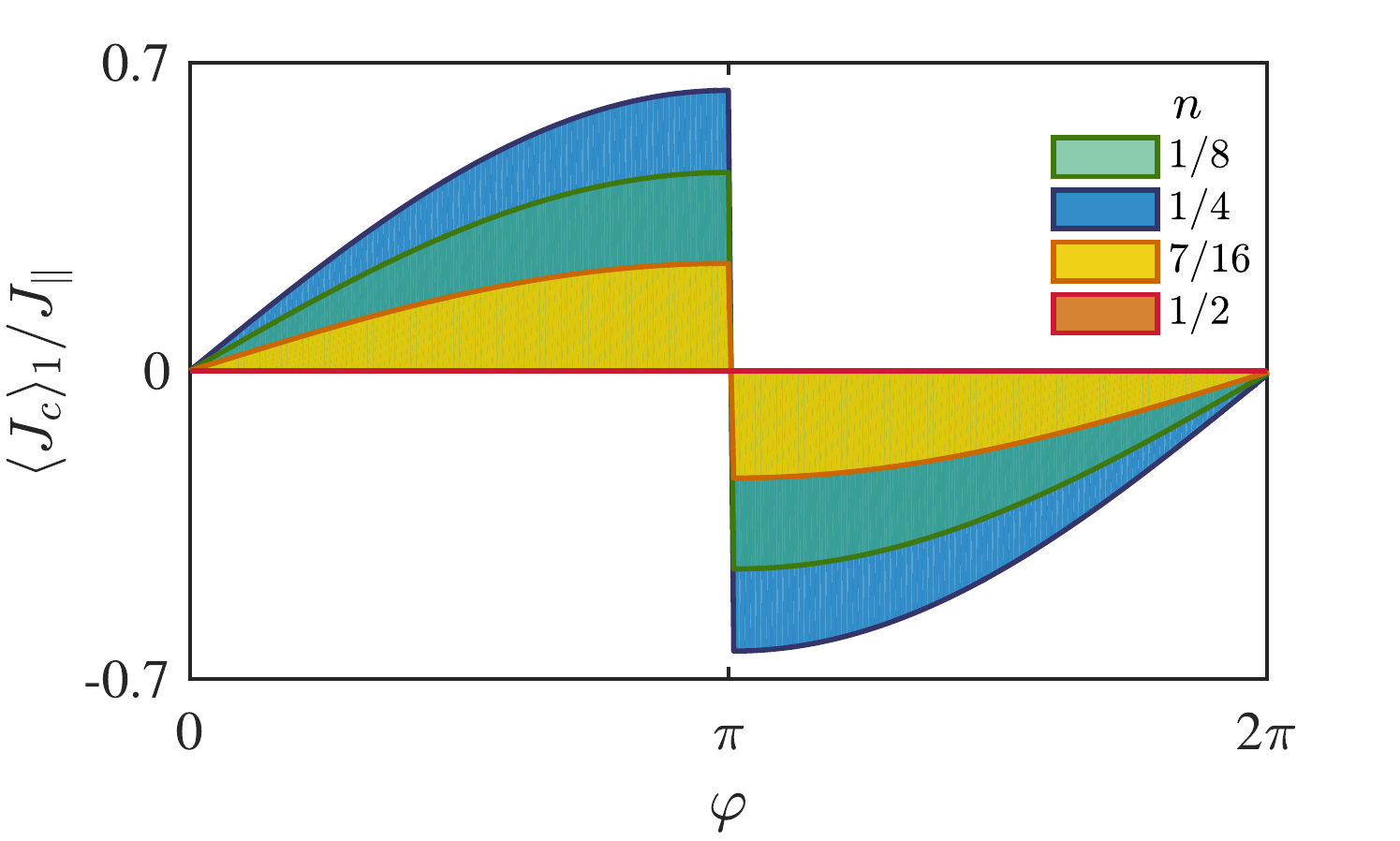}
 \caption{(Color online) Dependence of the chiral current on flux $\varphi$ and filling $n$ for very strong pump amplitude, $A\rightarrow\infty$ where only one self-consistent solution for the hopping amplitude exists. The current changes its direction at $\varphi=\pi$.}
\label{fig:Jc_diff_n_A_infty}
\end{figure}
\begin{figure}
\centering
\includegraphics[width=0.5\textwidth]{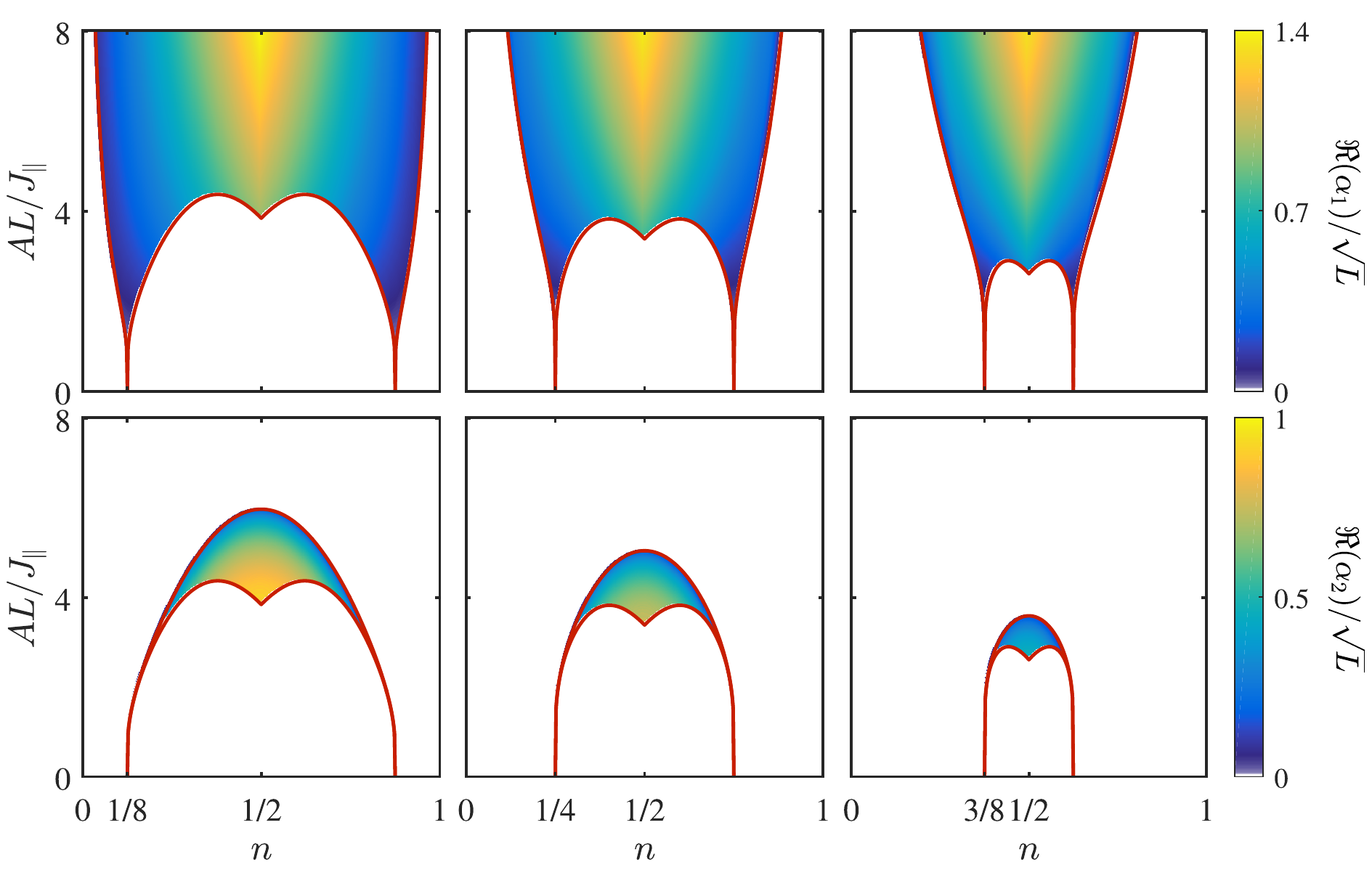}
\caption{(Color online)  Regime of existence of a self-organized chiral state for flux values $\varphi=\frac{\pi}{4},\frac{\pi}{2},\frac{3\pi}{4}$ from left to right.
 Upper panels: The red solid curves mark the critical values of pump strength $A_{cr}$ above which a self-consistent solution with a finite cavity field exists. Lower panels: the regime of existence of the second solution of scenario (iii) (lying in region c) is represented in between the lower and upper red solid line. In the self-organized phase the real part of the cavity field expectation value $\Re(\alpha)/\sqrt{L}$ is shown by the color code. Here, $\hbar \delta_{cp}= J_\parallel$ and $\hbar\kappa=0.05 J_\parallel$.}
\label{fig:alpha_A_n_diffPhi}
\end{figure}
\begin{figure}
\centering
\includegraphics[width=0.5\textwidth]{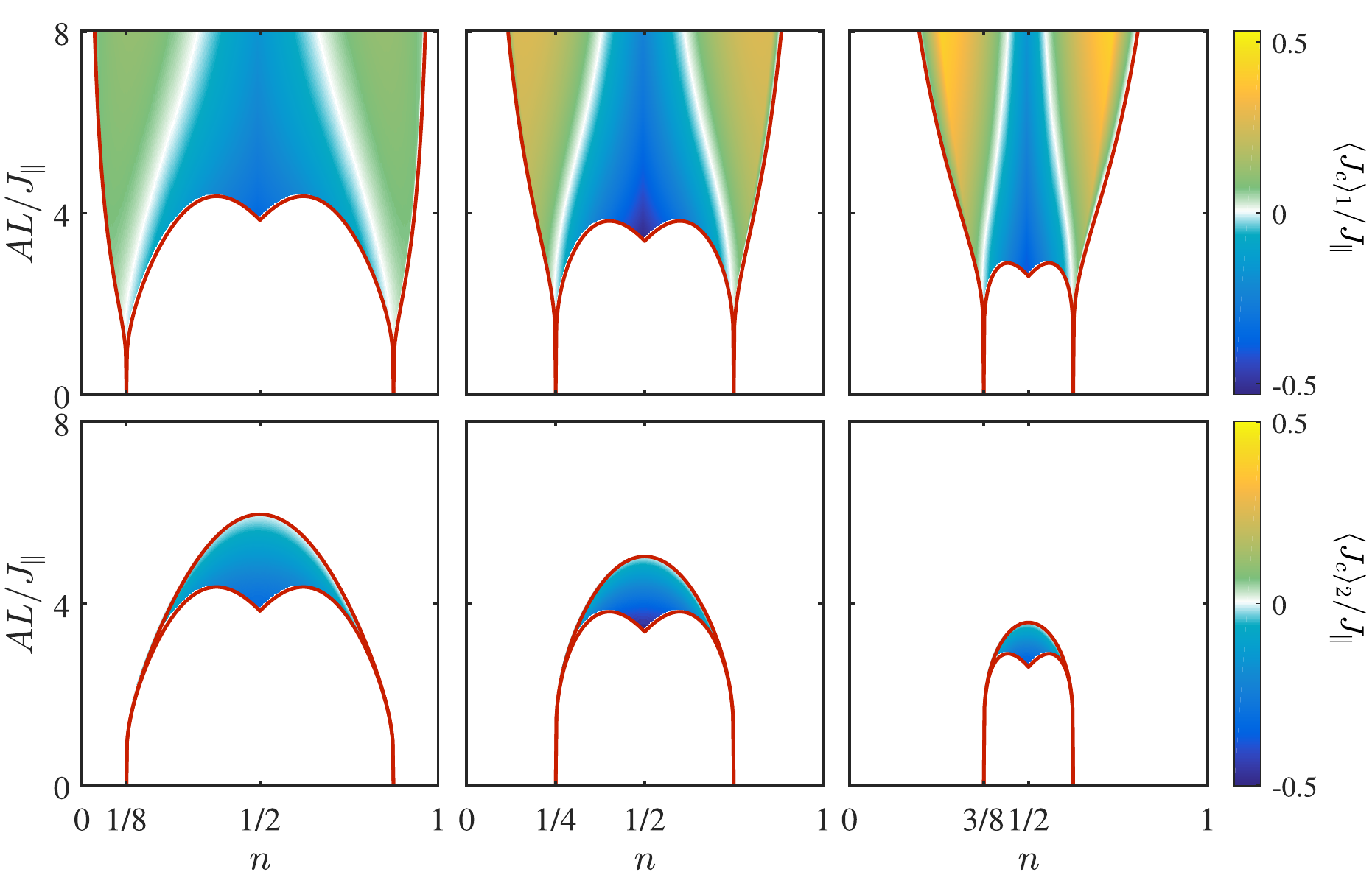}
\caption{(Color online)  Expectation value of the chiral current in the self-organized chiral state for $\varphi=\frac{\pi}{4},\frac{\pi}{2},\frac{3\pi}{4}$ from left to right versus filling $n$.
 Upper panels: The red solid curves mark the critical values of pump strength $A_{cr}$ above which a self-consistent solution with a finite cavity field exists. Lower panels: the regime of existence of the second solution of scenario (iii) (lying in region c) is represented in between the lower and upper red solid line. Parameters as in Fig.~\ref{fig:alpha_A_n_diffPhi}.}
\label{fig:Jc1_Jc2_A_n_diffPhi}
\end{figure}
In this subsection we discuss the properties of the self-organized states corresponding to the solutions found in the previous section. We show in Fig.~\ref{fig:alpha_phi_A_diffn} the real part of the expectation value of the cavity field $\Re(\alpha)$ which summarizes the regions in which a non-trivial self-consistent solution was found. Additionally, we plot in Fig.~\ref{fig:Jc1_phi_A_diffn} the value of the arising chiral current in this state. We represent the results as a function of the pump strength $A \propto \Omega_p^2$, which can be tuned in particular by the intensity of the transverse pump laser, and the flux $\varphi$, which can be adjusted by the lattice geometry and the wavelength of the pump laser as described in section \ref{sec:system}.    

The lower critical values $A_{cr}$ for the onset of a self-organized state with a finite cavity field are shown as solid (red) lines. A finite value of $A_{cr}$ is required for most values of the flux $\varphi$ and typically a sudden onset of the self-organized state with finite cavity occupation occurs. Only at the critical values $\varphi_{cr}$ (scenario (ii)), such self-organization arises for an infinitely small value of $A$ and persists for all values. One finds that a sudden jump to a finite value of the cavity field and of the chiral current arises at $A_{cr}$ for values of the flux $\varphi<\varphi_{cr} \le \pi$ (and $\varphi>2\pi-\varphi_{cr}$). This corresponds to the scenario (i) of the self-consistent solution. In contrast, for the range of flux $\varphi_{cr}<\varphi<2\pi-\varphi_{cr}$, the expectation value of the cavity field and, thus, of the expectation value of the chiral current vanishes and only slowly increases with increasing value of $A$. This corresponds to the scenario (iii) of the self-consistent solution. Note, that along the line $\varphi=\pi$ by symmetry reasons the chiral current vanishes and that another two curves of vanishing chiral current exist away of half filling. The behavior of the chiral current at fixed value of $A$ becomes more clear in the cuts shown in Fig.~\ref{fig:Jc12_phi_A}. The current shows depending on the value of $A$ very different dependence on the flux. For the first solution at low values of $A$, i.e.~$A=3.1 J_\parallel/L$ in Fig.~\ref{fig:Jc12_phi_A}, the current remains zero up to a critical value of the flux (intersection with the red line in Fig.~\ref{fig:Jc1_phi_A_diffn}). Subsequently, it first takes a negative value until it reaches its minimum, where it starts to grow to a positive value. This means that the chiral current inverts its direction. At a second critical value it vanishes again. Due to symmetry, the inverse dependence on the flux can be seen for values above $\varphi=\pi$. At intermediate values of $A$ (cf.~$A=15.9 J_\parallel/L$ in Fig.~\ref{fig:Jc12_phi_A}) the initial vanishing and negative regime of the current shrinks and an almost triangular shape is found. The inversion of the current only takes place at the symmetry point $\varphi=\pi$. 

When the pump strength is very large there exists only one self-consistent solution. For $A\to \infty$, the chiral current approaches a constant value $\langle J_c \rangle/J_\parallel=\frac{2}{\pi} \sin(\frac{\varphi}{2}) \sin(2\pi n)$. The direction of the current for values of $n<1/2$ and $\varphi<\pi$ is always the same and a change occurs at $\varphi=\pi$. The maximum value of chiral current for large pump strength occurs at quarter filling $n=\frac{1}{4}$ (and $n=\frac{3}{4}$) with $\varphi\to \pi$ and it goes to zero at half filling $n=\frac{1}{2}$ and very low or very high filling (Fig.~\ref{fig:Jc_diff_n_A_infty}).
 
Since for the flux $\varphi<\varphi_{cr}$ (and $\varphi>2\pi-\varphi_{cr}$) two solutions can exist, we show also the chiral current corresponding to the second solution. This solution only exists in a finite regime of values of $A$ as marked in the lower panel of Fig.~\ref{fig:Jc1_phi_A_diffn} and Fig.~\ref{fig:Jc12_phi_A}. 
As for the first solution, the chiral current corresponding to the second solution jumps to a finite value at the critical value of $A_{cr}$. However, for larger values of the pump strength $A$ the chiral current of the second solution decreases to zero, since it corresponds to the effective values of the ratio of the rung tunneling $J_\perp/J_\parallel$ which decrease with increasing value of $A$. At a fixed value of $A$ (Fig.~\ref{fig:Jc12_phi_A} lower panel) the second solution shows a rapid decrease of the current with increasing value of the flux ($\varphi<\pi$) to a minimal negative value followed by a rapid decrease to zero.

In Fig.~\ref{fig:alpha_A_n_diffPhi} and in Fig.~\ref{fig:Jc1_Jc2_A_n_diffPhi}, the real part of the expectation value of the cavity field and the chiral current is plotted versus the filling and the pump strength for different fluxes. The solid red line indicates the onset of a non-trivial solution of the self-consistency condition. The upper panels show the first solution. Since  $2\pi n=\varphi_{cr}$, the critical value of the flux is proportional to the filling. Additionally, at half filling the boundary of the regime of the self-consistent solution bends down and shows a cusp. Furthermore, the cavity field amplitude reaches a maximum at half filling. The chiral current shows a more complex behaviour. It is maximal close to the boundaries of the region where a non-trivial solution exists, and a line of vanishing current starts at the critical density and moves towards larger densities (for $n<1/2$). Across this line the chiral current changes sign and becomes maximally negative at half filling. 
Whereas the general form is similar for different values of the flux, the region of existence of the non-trivial solution shrinks for larger values of the flux and a stronger pump strength is needed at low densities. In contrast, the critical pump amplitude at half filling decreases slightly with increasing value of the flux. 

These results show that in most of the cases of existence the self-organized state of the fermionic atoms carries a chiral current. Depending on the filling this state is insulating or liquid in nature. Away from half-filling a chiral liquid is formed. In contrast, at half-filling one has to distinguish between two different solutions: for the first solution (region b) a chiral insulator arises, whereas the second solution corresponds to a chiral liquid (see Fig.~\ref{fig:Jc1_phi_A_diffn}). 

\section{Detection of the chiral current}
\label{sec:detection}
The self-organized chiral current can be measured in a very direct way by observing the superradiant scattering of a weak probe beam into an empty cavity mode. To this end, a magnetic field gradient is applied along the $y$-direction which leads to a potential offset $\Delta'$ between neighboring sites along the legs of the ladders. In addition, a weak probe beam with frequency $\omega'_{p}$ is applied along the $z$-direction. The frequency $\omega'_{p}$ is chosen such that a Raman process is induced between neighboring sites on a leg via the probe beam and an empty 'probe' cavity mode whose longitudinal mode number differs by two from the main cavity mode. 

The induced process can be described by the effective term
\begin{equation}
H_\mathrm{probe} = \hbar \tilde{\Omega}' \sum_{j,m=0,1}(-1)^m(b^\dag c_{m,j}^\dag c_{m,j+1}+\mathrm{H.c.})
\end{equation}
where $b$ denotes the annihilation operator of the probe cavity mode and $\tilde{\Omega}'$ is the two-photon Rabi frequency. The factor $(-1)^m $ takes into account that the spatial profile of the probe cavity mode has opposite sign on the two legs of the ladder. 

The above Hamiltonian can be written in terms of the directed tunneling $K_m = \frac{1}{L-1}\sum_j c_{m,j}^\dag c_{m,j+1}$ on leg $m$ and the chiral current as,
\begin{eqnarray}
&&\frac{H_\mathrm{probe}}{L-1}= \\
&&\frac{\hbar \tilde{\Omega}'}{2}\left[(b+b^{\dag})\sum_{m}(-1)^m (K_m + K_m^\dag)+i (b-b^{\dag})J_c/J_\parallel\right].\nonumber
\end{eqnarray}

The equations of motion for the probe cavity field are:
\begin{eqnarray}
-i \partial_t \langle b\rangle=&-&\frac{(L-1)\tilde{\Omega}'}{2}\left[(-1)^m\langle K_m + K_m^\dag \rangle
-i\langle J_c \rangle/J_\parallel\right]\nonumber \\ 
&-& ( \delta'_{cp}+ i \kappa' ) \langle b\rangle\nonumber\\
-i \partial_t \langle b^\dag\rangle=&+& \frac{(L-1)\tilde{\Omega}'}{2}\left[(-1)^m\langle K_m + K_m^\dag \rangle+i\langle J_c \rangle/J_\parallel\right]\nonumber\\
&+&( \delta'_{cp}- i \kappa' ) \langle b^\dag\rangle
\end{eqnarray}

where $\delta'_{cp}=\omega'_{c}-\omega'_{p}+\Delta'/\hbar$, $\omega'_{c}$ denotes the frequency and $\kappa'$ the decay rate of the probe cavity mode. 
In the stationary state, the chiral current is directly mapped onto the probe cavity field as

\begin{eqnarray}
\langle Jc \rangle/=\frac{J_\parallel}{(L-1)\tilde{\Omega'}}  \left(i \delta'_{cp}  \langle -b +b^\dag\rangle+ \kappa'\langle b+b^\dag\rangle \right).
\end{eqnarray}

Experimentally, 
the chiral current can thus be directly measured by observing the appropriate quadrature 
using a heterodyne detection scheme.
\section{Conclusion}
In this work we have investigated the steady-state diagram of a coupled atom-cavity system on a ladder geometry. The coupling is realized via a Raman process employing  the cavity field and a transverse running wave pump beam. This induces a cavity-assisted tunneling process along the rungs which comprises a spatially dependent phase imprint. Above a critical pump strength (which can be zero) we have found a spontaneous self-organization of the system into a state in which the emergent cavity field induces a strong artificial magnetic field for the atoms. In this artificial magnetic field, the atoms aquire a chiral current and the arising state is typically a chiral insulator for certain regimes at half filling or a chiral liquid. Only narrow lines along which the chiral current vanishes exist in the phase diagram. The occupation of the cavity field can either take place via a sudden jump at a critical value of the pump strength or via a slow activation. 

Beyond the mean-field description, an effective dissipative dynamics with jump operator $K_\perp$ and rate $\Gamma \sim \frac{\kappa \tilde{\Omega}^2}{\delta_{cp}^2+ \kappa^2}$ could drive the atomic system away from the ground state determined by $H_F$ into a steady-state which is a dynamical equilibrium between driving and damping \cite{RitschEsslinger2013}. For the running-wave pump configuration considered above, this could result on long time-scales in a transfer of the entire atomic population into the right leg of the ladders (see Fig.~\ref{fig:cavity}). This can be avoided by adding a second running-wave pump laser field along the $y$-direction which together with a second cavity mode (separated from the first cavity mode by twice the free spectral range) drives Raman transitions along the rungs into the opposite direction \cite{KollathBrennecke2016}. 

The presented work enables the realization of topologically non-trivial phases as attractor states of a dissipative dynamics. 
Additionally, we detail how the emerging chiral current can be measured experimentally in a direct and non-destructive way using the cavity output field. 

The exact characterization of the dissipative temporal dynamics going beyond the characterization of the steady-state phase diagram is of great interest for further studies. Additionally, the extension of the presented scheme into two dimensions, where true edge states separated by a bulk exist, is a direction to explore. Such edge states could have a protection by the dissipative attractor dynamics and by their topological nature. 

We thank M.~Fleischhauer,  H.~Monien, F.~Piazza, H.~Ritsch, S.~Wolff and W.~Zwerger for fruitful discussion. We acknowledge financial support from the DFG and the ERC (Grant Number 648166). 
\appendix
\section{Derivation of the effective Hamiltonian}
\label{app:derivation}

A fermionic quantum gas placed in an optical cavity and subjected to optical lattice potentials and a transversal pump beam can be described in the rotating wave approximation by the Hamiltonian \cite{RitschEsslinger2013}
\beq
H\approx H_{g}+H_e+H_{c}+H_{ac}+H_{ap},
\eeq
where we define the different terms in the following.
We assume that only two internal states of the atom, the ground and one excited state, are important for the atomic dynamics. The atomic motion is described by the first contribution
\begin{multline}
H_{g}=\int\mathrm{d}^3 {\bf r}\; \left[ \Psi_g^{\dag}({\bf r}) \left(
-\frac{\hbar^2}{2m}\nabla^2+V_g({\bf r})\right)\Psi_g({\bf r})\right] \nonumber\\ 
H_{e}=\int\mathrm{d}^3 {\bf r}\;  \Psi_e^{\dag}({\bf r}) \left[\left( -\frac{\hbar^2}{2m}\nabla^2+\hbar\omega_{ep}+V_e({\bf r})\right)\Psi_e({\bf r})\right],
\nonumber
\end{multline}  
where $\Psi_g({\bf r})$ and $\Psi_e({\bf r})$ denote the fermionic annihilation operators at position ${\bf r}$ in the
ground state and the excited state, respectively. The excited state operator is defined in the frame rotating at the pump frequency. The field operators obey the usual fermionic anti-commutation relations
\begin{subequations}
\begin{align}
\left\{ \Psi_f({\bf r}),\Psi_{f^\prime}^\dag({\bf r}^\prime)\right\} & =
\delta \left({\bf r}-{\bf r}^\prime\right)\delta_{f,f^\prime}\\ 
\left\{\Psi_f({\bf r}),\Psi_{f^\prime}({\bf r}^\prime)\right\} & =\left\{
\Psi_f^\dag({\bf r}),\Psi_{f^\prime}^\dag({\bf r}^\prime)\right\}=0,
\end{align}
\end{subequations}
where $f,f^\prime\in\{e,g\}$. 
The atomic frequency between the ground and the excited state is given by $\omega_{e}$ and the detuning of the pump laser from the atomic transition is defined by $\omega_{ep}=\omega_{e}-\omega_p$. The potentials $V_e({\bf r})$ and $V_g({\bf r})$ are the external potentials for the atom in the excited and the ground state, respectively. These contain the optical lattice potential and other possible trapping potentials. The interaction between the atoms in the ground and the excited state has been neglected, since the excited state is barely populated for large detuning $\omega_{ep}$. 

The second term describes the cavity field dynamics
\begin{equation*}
H_{c}=\hbar\omega_{cp} a^\dag a.
\end{equation*}
 Here, $\omega_{cp}=\omega_c-\omega_p$ is the detuning between the dispersively shifted resonance frequency $\omega_c$ of the cavity mode and the pump frequency $\omega_p$ and $a$ the annihilation operator of cavity photons in the frame rotating at $\omega_p$. In addition to the unitary evolution described by the Hamiltonian, the cavity field is subjected to loss which require the description by a Lindblad master equation.

The coupling between the atoms and the cavity field is represented by
\beq 
H_{ac}=\hbar g_0 \int\mathrm{d}^3 {\bf r} \; \left(\Psi_g^\dag({\bf r}) \cos({\bf k}_c.{\bf r}) a^\dagger \Psi_e({\bf r}) + \mathrm{h.c.}\right),
\eeq 
where $g_0$ is the vacuum-Rabi frequency of the cavity and ${\bf k}_c$ is the wave vector of the cavity mode.

The interaction with the pump laser beam, which coherently
drives the atoms, reads 
\beq 
H_{ap}=\hbar\Omega_p\int\mathrm{d}^3 {\bf r}\; \left(\Psi_g^\dag({\bf r}) e^{-i{\bf k}_p.{\bf r}}  \Psi_e({\bf r}) +  \mathrm{h.c.} \right),
\eeq
where $\Omega_p$ denotes the Rabi frequency of the pump beam. 
Since the internal time-scales are fast and the excited state is hardly occupied for far off-resonant driving, we can adiabatically eliminate the excited state in order to obtain an effective description of the dynamics of the atomic ground state and the cavity field. 
Using the equation of motion of the excited state
\begin{multline}\label{eq:psi_e}
i\hbar\frac{\partial \Psi_e({\bf r})}{\partial t} =
\left[-\frac{\hbar^2}{2m}\nabla^2+V_e({\bf r})+\hbar\omega_{ep}\right]\Psi_e({\bf r})\\
+ \left[ \hbar g_0 \cos({\bf k}_c.{\bf r}) a + \hbar\Omega_p e^{i{\bf k}_p.{\bf r}} \right]\Psi_g({\bf r}),
\end{multline}
its stationary value is found to be 
\beq 
\Psi_e({\bf r})=-\frac{1}{\omega_{ep}}\left[ g_0 \cos({\bf k}_c.{\bf r}) a + \Omega_p e^{i{\bf k}_p.{\bf r}} \right]\Psi_g({\bf r}). 
\eeq

The equations of motion for the atomic ground state and the cavity field which result from substituting the stationary value of the excited field 
can be obtained from the following effective Hamiltonian
\begin{align}\label{ad_ham}
H_{\textrm{eff}}=& H_{c} + H_{g} + H_{ac}\nonumber\\
H_{ac}=&-\frac{\hbar g_0\Omega_p}{\omega_{ep}}\int\mathrm{d}^3{\bf r}\left( e^{i{\bf k}_p.{\bf r}} a^\dag + e^{-i{\bf k}_p.{\bf r}} a \right)\nonumber\\
&\times \cos({\bf k}_c.{\bf r})\Psi_g^\dag({\bf r})\Psi_g({\bf r})\nonumber\\
\end{align}
combined with the dissipative term of the Lindblad equation for the cavity losses. 
Here we have only taken into account the two-photon transitions involving one pump and one cavity photon which will lead to the cavity-induced tunneling, and we have neglected the AC-Stark shift induced by intra-cavity photons or by the pump beam. 

In a sufficiently strong optical lattice potential, a convenient choice is to expand the fermionic field operators into the corresponding Wannier basis of the lattice
\begin{align}
\Psi_g^\dag({\bf r})=\sum_{m,j} w^*({\bf r-R}_{m,j})c_{m,j}^\dag,
\end{align}
where ${\bf R}_{m,j}$ denotes the position of the lattice site $j$ on leg $m$ and $c_{m,j}^\dag$ represents the corresponding creation operator of the fermionic state on leg $m$ and site $j$. The advantage of such a representation is the localization of the Wannier functions in the lattice wells. 

Using the expansion into Wannier functions and neglecting off-resonant terms of the two-photon transition, the resulting effective Hamiltonian $H_F$ is given by 
equation (\ref{eq:Heff}) in the main text. 
The factors  $\phi_\parallel$ and $\phi_{\perp}$ are effective parameters which can be related to the microscopic parameters of the underlying geometry.  
In particular, along the $y$-direction the onsite contribution of the overlap integrals of the Wannier functions is typically most important such that $\phi_\parallel$ is dominated by 
$$
\phi_{\parallel,0}(k_p)= \int\mathrm{d}y \; w^*(y) w(y)e^{-ik_p y}.
$$

Along the $x$-direction two different processes can give important contributions depending on the chosen lattice geometry. 
The first one stems from the overlap between the Wannier functions of neighbouring lattice wells and is given by 
$$
 \phi_{\perp,\pm}(k_c)=\int\mathrm{d}x \; w^*(x) w(x\pm d_\perp) \cos(k_c x).
$$
Here $d_\perp$ is the lattice spacing along the rungs of the ladder.
The second contribution stems from the oscillating energy offset between the two sites on a rung. 
The amplitude of the energy offset is related to the on-site overlap integrals 
$$
\phi_{\perp,m}(k_c)=\int\mathrm{d}x \; w^*(x-md_\perp) w(x-md_\perp) \cos\left(k_c (x + m d_\perp) \right),
$$
 which are distinct on the two different legs $m=0,1$.
The coupling to the cavity mode induces a time-modulation of the potential offset of the two sites along a rung and by this leads to an effective tunneling with an amplitude proportional to the difference $\phi_{\perp,1}-\phi_{\perp,0}$ and inverse proportional to the oscillation frequency.
We assume that both parts are included in the effective parameter $\phi_\perp$ of the main text. 


\begin{thebibliography}{10}
\bibitem{Chu1998}
S.~Chu,
\newblock Rev. Mod. Phys. {\bf 70}, 685 (1998).

\bibitem{CohenTannoudji1998}
C.~N. Cohen-Tannoudji,
\newblock Rev. Mod. Phys. {\bf 70}, 707 (1998).

\bibitem{Phillips1998}
W.~D. Phillips,
\newblock Rev. Mod. Phys. {\bf 70}, 721 (1998).

\bibitem{Foot2005}
C.~Foot,
\newblock {\em Atomic physics} (Oxford University Press, 2005).

\bibitem{PethickSmith}
C.~J. Pethick and H.~Smith,
\newblock {\em {B}ose-{E}instein Condensation in Dilute Gases} (Cambridge
  University Press, 2002).

\bibitem{RitschEsslinger2013}
H.~Ritsch, P.~Domokos, F.~Brennecke, and T.~Esslinger,
\newblock Rev. Mod. Phys. {\bf 85}, 553 (2013).

\bibitem{Dicke1954}
R.~H. Dicke,
\newblock Phys. Rev. {\bf 93}, 99 (1954).

\bibitem{HeppLieb1973}
K.~Hepp and E.~H. Lieb,
\newblock Annals of Physics {\bf 76}, 360  (1973).

\bibitem{WangHioe1973}
Y.~K. Wang and F.~T. Hioe,
\newblock Phys. Rev. A {\bf 7}, 831 (1973).

\bibitem{DomokosRitsch2002}
P.~Domokos and H.~Ritsch,
\newblock Phys. Rev. Lett. {\bf 89}, 253003 (2002).

\bibitem{NagyDomokos2008}
G.~S. D.~Nagy and P.~Domokos,
\newblock Eur.~Phys.~J.~D {\bf 48}, 127 (2008).

\bibitem{BaumannEsslinger2010}
K.~Baumann, C.~Guerlin, F.~Brennecke, and T.~Esslinger,
\newblock Nature {\bf 464}, 1301 (2010).

\bibitem{KlinderHemmerich2015}
J.~Klinder, H.~Keßler, M.~Wolke, L.~Mathey, and A.~Hemmerich,
\newblock Proceedings of the National Academy of Sciences {\bf 112}, 3290
  (2015), http://www.pnas.org/content/112/11/3290.full.pdf.

\bibitem{PiazzaZwerger2013}
F.~Piazza, P.~Strack, and W.~Zwerger,
\newblock arXiv:1305.2928  (2013).

\bibitem{DimerCarmichael2007}
F.~Dimer, B.~Estienne, A.~S. Parkins, and H.~J. Carmichael,
\newblock Phys. Rev. A {\bf 75}, 013804 (2007).

\bibitem{BadenBarrett2014}
M.~P. Baden, K.~J. Arnold, A.~L. Grimsmo, S.~Parkins, and M.~D. Barrett,
\newblock Phys. Rev. Lett. {\bf 113}, 020408 (2014).

\bibitem{BhaseenKeeling2012}
M.~J. Bhaseen, J.~Mayoh, B.~D. Simons, and J.~Keeling,
\newblock Phys. Rev. A {\bf 85}, 013817 (2012).

\bibitem{LiuJia2011}
N.~Liu {\em et~al.},
\newblock Phys. Rev. A {\bf 83}, 033601 (2011).

\bibitem{KulkarniTuereci2013}
M.~Kulkarni, B.~\"Oztop, and H.~E. T\"ureci,
\newblock Phys. Rev. Lett. {\bf 111}, 220408 (2013).

\bibitem{KonyaDomokos2014}
G.~K\'onya, G.~Szirmai, and P.~Domokos,
\newblock Phys. Rev. A {\bf 90}, 013623 (2014).

\bibitem{PiazzaRitsch2015}
F.~Piazza and H.~Ritsch,
\newblock Phys. Rev. Lett. {\bf 115}, 163601 (2015).

\bibitem{SchuetzMorigi2014}
S.~Sch\"utz and G.~Morigi,
\newblock Phys. Rev. Lett. {\bf 113}, 203002 (2014).

\bibitem{KlinderHemmerich2015b}
J.~Klinder, H.~Ke\ss{}ler, M.~R. Bakhtiari, M.~Thorwart, and A.~Hemmerich,
\newblock Phys. Rev. Lett. {\bf 115}, 230403 (2015).

\bibitem{LandingEsslinger2015}
R.~Landig {\em et~al.},
\newblock arXiv:1511.00007  (2015).

\bibitem{LarsonLewenstein2008}
J.~Larson, B.~Damski, G.~Morigi, and M.~Lewenstein,
\newblock Phys. Rev. Lett. {\bf 100}, 050401 (2008).

\bibitem{MaschlerRitsch2005}
C.~Maschler and H.~Ritsch,
\newblock Phys. Rev. Lett. {\bf 95}, 260401 (2005).

\bibitem{MaschlerRitsch2008}
C.~Maschler, I.~B. Mekhov, and H.~Ritsch,
\newblock The European Physical Journal D {\bf 46}, 545 (2008).

\bibitem{NiedenzuRitsch2010}
W.~Niedenzu, R.~Schulze, A.~Vukics, and H.~Ritsch,
\newblock Phys. Rev. A {\bf 82}, 043605 (2010).

\bibitem{SilverSimons2010}
A.~O. Silver, M.~Hohenadler, M.~J. Bhaseen, and B.~D. Simons,
\newblock Phys. Rev. A {\bf 81}, 023617 (2010).

\bibitem{VidalMorigi2010}
S.~Fern\'andez-Vidal, G.~De~Chiara, J.~Larson, and G.~Morigi,
\newblock Phys. Rev. A {\bf 81}, 043407 (2010).

\bibitem{LiHofstetter2013}
Y.~Li, L.~He, and W.~Hofstetter,
\newblock Phys. Rev. A {\bf 87}, 051604 (2013).

\bibitem{BakhtiariThorwart2015}
M.~R. Bakhtiari, A.~Hemmerich, H.~Ritsch, and M.~Thorwart,
\newblock Phys. Rev. Lett. {\bf 114}, 123601 (2015).

\bibitem{SafaeiGremaud2015}
S.~Safaei, C.~Miniatura, and B.~Gr\'emaud,
\newblock Phys. Rev. A {\bf 92}, 043810 (2015).

\bibitem{LarsonLewenstein2008b}
J.~Larson, G.~Morigi, and M.~Lewenstein,
\newblock Phys. Rev. A {\bf 78}, 023815 (2008).

\bibitem{MuellerSachdev2012}
M.~M\"uller, P.~Strack, and S.~Sachdev,
\newblock Phys. Rev. A {\bf 86}, 023604 (2012).

\bibitem{PiazzaStrack2014}
F.~Piazza and P.~Strack,
\newblock Phys. Rev. Lett. {\bf 112}, 143003 (2014).

\bibitem{KeelingSimons2014}
J.~Keeling, J.~Bhaseen, M., and D.~Simons, B.,
\newblock Phys. Rev. Lett. {\bf 112}, 143002 (2014).

\bibitem{ChenZhai2014}
Y.~Chen, Z.~Yu, and H.~Zhai,
\newblock Phys. Rev. Lett. {\bf 112}, 143004 (2014).

\bibitem{GopalakrishnanGoldbart2009}
S.~Gopalakrishnan, B.~L. Lev, and P.~M. Goldbart,
\newblock Nat.~Phys. {\bf 5}, 845 (2009).

\bibitem{NimmrichterArndt2010}
S.~Nimmrichter, K.~Hammerer, P.~Asenbaum, H.~Ritsch, and M.~Arndt,
\newblock New Journal of Physics {\bf 12}, 083003 (2010).

\bibitem{StrackSachdev2011}
P.~Strack and S.~Sachdev,
\newblock Phys. Rev. Lett. {\bf 107}, 277202 (2011).

\bibitem{GopalakrishnanGoldbart2011}
S.~Gopalakrishnan, B.~L. Lev, and P.~M. Goldbart,
\newblock Phys. Rev. Lett. {\bf 107}, 277201 (2011).

\bibitem{HabibianMorigi2013}
H.~Habibian, A.~Winter, S.~Paganelli, H.~Rieger, and G.~Morigi,
\newblock Phys. Rev. Lett. {\bf 110}, 075304 (2013).

\bibitem{JanotRosenow2013}
A.~Janot, T.~Hyart, P.~R. Eastham, and B.~Rosenow,
\newblock Phys. Rev. Lett. {\bf 111}, 230403 (2013).

\bibitem{BuchholdDiehl2013}
M.~Buchhold, P.~Strack, S.~Sachdev, and S.~Diehl,
\newblock Phys. Rev. A {\bf 87}, 063622 (2013).

\bibitem{DengYi2014}
Y.~Deng, J.~Cheng, H.~Jing, and S.~Yi,
\newblock Phys. Rev. Lett. {\bf 112}, 143007 (2014).

\bibitem{DongPu2014}
L.~Dong, L.~Zhou, B.~Wu, B.~Ramachandhran, and H.~Pu,
\newblock Phys. Rev. A {\bf 89}, 011602 (2014).

\bibitem{PanGuo2015}
J.-S. Pan, X.-J. Liu, W.~Zhang, W.~Yi, and G.-C. Guo,
\newblock Phys. Rev. Lett. {\bf 115}, 045303 (2015).

\bibitem{PadhiGosh2014}
B.~Padhi and S.~Ghosh,
\newblock Phys. Rev. A {\bf 90}, 023627 (2014).

\bibitem{MivehvarFeder2014}
F.~Mivehvar and D.~L. Feder,
\newblock Phys. Rev. A {\bf 89}, 013803 (2014).

\bibitem{MivehvarFeder2015}
F.~Mivehvar and D.~L. Feder,
\newblock Phys. Rev. A {\bf 92}, 023611 (2015).

\bibitem{MekhovRitsch2007}
I.~B. Mekhov, C.~Maschler, and H.~Ritsch,
\newblock Phys. Rev. Lett. {\bf 98}, 100402 (2007).

\bibitem{ChenMeystre2007}
W.~Chen, D.~Meiser, and P.~Meystre,
\newblock Phys. Rev. A {\bf 75}, 023812 (2007).

\bibitem{ChenMeystre2009}
W.~Chen and P.~Meystre,
\newblock Phys. Rev. A {\bf 79}, 043801 (2009).

\bibitem{BhattacherjeeMan2010}
A.~Bhattacherjee, T.~Kumar, and M.~Mohan,
\newblock Central European Journal of Physics {\bf 8}, 850 (2010).

\bibitem{MekhovRitsch2012}
I.~B. Mekhov and H.~Ritsch,
\newblock Journal of Physics B: Atomic, Molecular and Optical Physics {\bf 45},
  102001 (2012).

\bibitem{OeztopTuereci2012}
B.~\"Oztop, M.~Bordyuh, O.~E. M\"ustecaplıoğlu, and H.~E. Tr"ueci,
\newblock New Journal of Physics {\bf 14}, 085011 (2012).

\bibitem{KozlowskiMekhov2015}
W.~Kozlowski, S.~F. Caballero-Benitez, and I.~B. Mekhov,
\newblock Phys. Rev. A {\bf 92}, 013613 (2015).

\bibitem{Brennecke2013}
F.~Brennecke {\em et~al.},
\newblock Proceedings of the National Academy of Sciences {\bf 110}, 11763
  (2013).

\bibitem{Landig2015}
R.~Landig, F.~Brennecke, R.~Mottl, T.~Donner, and T.~Esslinger,
\newblock Nature communications {\bf 6} (2015).

\bibitem{PanYi2015}
J.-S. Pan, X.-J. Liu, W.~Zhang, and W.~Yi,
\newblock arXiv:1509.02993  (2015).

\bibitem{KollathBrennecke2016}
C.~Kollath, A.~Sheikhan, S.~Wolff, and F.~Brennecke,
\newblock arXiv:1502.01817  (2015).

\bibitem{HasanKane2010}
M.~Z. Hasan and C.~L. Kane,
\newblock Rev. Mod. Phys. {\bf 82}, 3045 (2010).

\bibitem{SternLindner2013}
A.~Stern and N.~H. Lindner,
\newblock Science {\bf 339}, 1179 (2013),
  http://www.sciencemag.org/content/339/6124/1179.full.pdf.

\bibitem{DalibardOehberg2011}
J.~Dalibard, F.~Gerbier, G.~Juzeli\ifmmode~\bar{u}\else \={u}\fi{}nas, and
  P.~\"Ohberg,
\newblock Rev. Mod. Phys. {\bf 83}, 1523 (2011).

\bibitem{JakschZoller2003}
D.~Jaksch and P.~Zoller,
\newblock New Journal of Physics {\bf 5}, 56 (2003).

\bibitem{AidelsburgerBloch2013}
M.~Aidelsburger {\em et~al.},
\newblock Phys. Rev. Lett. {\bf 111}, 185301 (2013).

\bibitem{MiyakeKetterle2013}
H.~Miyake, G.~A. Siviloglou, C.~J. Kennedy, W.~C. Burton, and W.~Ketterle,
\newblock Phys. Rev. Lett. {\bf 111}, 199903 (2013).

\bibitem{AidelsburgerGoldman2014}
M.~Aidelsburger {\em et~al.},
\newblock arXiv:1407.4205  (2014).

\bibitem{AtalaBloch2014}
M.~Atala {\em et~al.},
\newblock Nat.~Phys. {\bf 10}, 588 (2014).

\bibitem{JotsuEsslinger2014}
G.~Jotzu {\em et~al.},
\newblock Nature , 237 (2014).

\bibitem{CarrNersesyan2006}
S.~T. Carr, B.~N. Narozhny, and A.~A. Nersesyan,
\newblock Phys. Rev. B {\bf 73}, 195114 (2006).

\bibitem{RouxPoilblanc2007}
G.~Roux, E.~Orignac, S.~White, and D.~Poilblanc,
\newblock Phys. Rev. B {\bf 76}, 195105 (2007).

\bibitem{JaefariFradkin2012}
A.~Jaefari and E.~Fradkin,
\newblock Phys. Rev. B {\bf 85}, 035104 (2012).

\bibitem{HuegelParedes2014}
D.~H\"ugel and B.~Paredes,
\newblock Phys. Rev. A {\bf 89}, 023619 (2014).

\bibitem{TokunoGeorges2014}
A.~Tokuno and A.~Georges,
\newblock New Journal of Physics {\bf 16}, 073005 (2014).

\end{thebibliography}
\end{document}